\documentclass[aps,reprint,showpacs,floatfix,superscriptaddress,amsmath,amssymb,prb]{revtex4-1}
\usepackage{graphicx}
\usepackage{hyperref}
\usepackage{multirow}

\newcommand{\ket} [1] {| #1 \rangle}
\newcommand{\bra} [1] {\langle #1 |}
\newcommand{\braket}[2]{\langle #1 | #2 \rangle}

\newcommand{\veclat}[1]{\mathbb{V}^{(\mathcal{#1})}}

\newcommand{\eref}[1]{Eq.~(\ref{#1})}
\newcommand{\eeref}[1]{(\ref{#1})}

\newcommand{\fref}[1]{Fig.~\ref{#1}}
\newcommand{\fuse}[3]{#1#2\!\rightarrow\!#3}

\usepackage{array}

\begin{document}
\title{Matrix product states for anyonic systems and efficient simulation of dynamics}
\author{Sukhwinder Singh}
\affiliation{Center for Engineered Quantum Systems; Dept. of Physics \& Astronomy, \\Macquarie University, 2109 NSW, Australia}
\author{Robert N. C. Pfeifer}
\affiliation{Perimeter Institute for Theoretical Physics, Waterloo, Ontario, N2L 2Y5, Canada}
\author{Guifre Vidal}
\affiliation{Perimeter Institute for Theoretical Physics, Waterloo, Ontario, N2L 2Y5, Canada}
\author{Gavin K. Brennen}
\affiliation{Center for Engineered Quantum Systems; Dept. of Physics \& Astronomy, \\Macquarie University, 2109 NSW, Australia}

\begin{abstract}
Matrix product states (MPS) have proven to be a very successful tool to study lattice systems with local degrees of freedom such as spins or bosons. 
Topologically ordered systems can support anyonic particles which are labeled by conserved topological charges and collectively carry non-local degrees of freedom. 
 In this paper we extend the formalism of MPS to lattice systems of anyons. The anyonic MPS is constructed from tensors that explicitly conserve topological charge. We describe how to adapt the  time-evolving block decimation (TEBD) algorithm to the anyonic MPS in order to simulate dynamics under a local and charge-conserving Hamiltonian. To demonstrate the effectiveness of anyonic TEBD algorithm, we used it to simulate $(i)$ the ground state (using imaginary time evolution) of an infinite 1D critical system of (a) Ising anyons and (b) Fibonacci anyons both of which are well studied, and $(ii)$ the real time dynamics of an anyonic Hubbard-like model of a single Ising anyon hopping on a ladder geometry with an anyonic flux threading each island of the ladder. Our results pertaining to $(ii)$ give insight into the transport properties of anyons. The anyonic MPS formalism can be readily adapted to study systems with conserved symmetry charges, as this is equivalent to a specialization of the more general anyonic case.
\end{abstract}


\maketitle

\section{Introduction\label{sec:intro}}
Anyons are exotic quasiparticles that exhibit non-trivial exchange statistics and arise as low lying excitations of topological phases of matter. There is a promising experimental program to observe anyons in condensed matter systems such as Fractional Quantum Hall systems, Majorana edge modes of nanowires, and two dimensional spin liquids.\cite{Pachos} From a theoretical viewpoint, many-body systems of anyons offer a realm of new physics to explore. For example, just as chains of interacting integer spin systems describe different physics from half integer systems,\cite{Haldane} interacting chains of anyons exhibit properties that depend on the topological charges and braiding and fusion rules of the corresponding anyon model.\cite{TTWL} Subsequently, several paradigmatic lattice models of interacting anyons have been proposed and studied to gain insight into the many-body physics of these particles.\cite{Feiguin07,FRBM,Poilblanc11,Gils13,Poilblanc13} 

The framework of Matrix Product States \cite{Fannes92,White92, Ostlund95,Vidal03,Vidal04,PerezGarcia07} has played an instrumental role in the study of lattice spin systems in recent decades, especially for systems in one spatial dimension. In particular, the MPS forms the basis of two highly successful simulation algorithms, namely, the Density Matrix Renormalization Group (DMRG) algorithm \cite{White92} and the Time-evolving Block Decimation (TEBD) algorithm.\cite{Vidal03,Vidal04} The latter, along with its variations (often collectively referred to as time-dependent DMRG \cite{Daley04,White04,Schollwock05,Shi06,Vidal071}), allows for efficient simulation of time evolution of lattice systems made of hundreds of sites, and also of systems with infinite size \cite{Vidal071} in the presence of translation invariance. In this paper we generalize the MPS formalism and the TEBD algorithm for lattice systems of anyons. 
 
\begin{figure}
  \includegraphics[width=8cm]{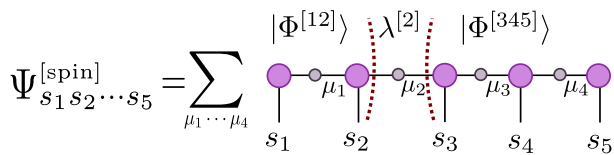}
\caption{MPS decomposition of a pure state $\ket{\Psi^{\mbox{\tiny [spin]}}}$ of a lattice of $L=5$ spins $\{s_1,s_2,\ldots,s_5\}$, as described in Ref.~\onlinecite{Vidal03}. An open index $s_i$ labels a basis for the spin at site $i$. A bond index $\mu_i$ is associated with a Schmidt decomposition \eeref{eq:spinsd} of state $\ket{\Psi^{\mbox{\tiny [spin]}}}$. For example, here we indicate the Schmidt decomposition for the bipartition $[12]:[345]$ where $\lambda^{^{[2]}}$ is a diagonal matrix with diagonal entries $\lambda^{^{{[2]}}}_{\mu_2}\geq 0$, and $\ket{\Phi^{^{[12]}}_{\mu_2}}$ and $\ket{\Phi^{^{[345]}}_{\mu_2}}$ are the Schmidt bases for the two parts as described by the tensors to the left and to the right of index $\mu_2$ respectively.\label{fig:intro1}}
\end{figure}

A pure state $\ket{\Psi^{\mbox{\tiny [spin]}}}$ of a lattice made of $L$ spins $\{s_1,s_2,\ldots s_L\}$ can be expanded in a tensor product basis,
\begin{equation}
\ket{\Psi^{\mbox{\tiny [spin]}}} = \sum_{s_1s_2\ldots s_L} \Psi^{\mbox{\tiny [spin]}}_{s_1s_2\ldots s_L} \ket{s_1}\otimes\ket{s_2}\otimes\ldots\ket{s_L},\label{eq:spinstate}
\end{equation}
where $\Psi^{\mbox{\tiny [spin]}}_{s_1s_2\ldots s_L}$ are complex coefficients and $\{\ket{s_i}\}$ denotes the local basis for site $i$. As described in Ref.~\onlinecite{Vidal03}, the MPS decomposition of $\ket{\Psi^{\mbox{\tiny [spin]}}}$ corresponds to a decomposition of tensor $\Psi^{\mbox{\tiny [spin]}}_{s_1s_2\ldots s_L}$ into a network of tensors that are interconnected as shown in \fref{fig:intro1}. An open or \textit{physical} index $s_i$ in the MPS labels the local basis $\ket{s_i}$ for the spin at site $i$. A \textit{bond} index $\mu_i$ corresponds to the Schmidt decomposition (see Appendix \ref{sec:appAA}) of state $\ket{\Psi^{\mbox{\tiny [spin]}}}$, given according to the bipartition $[1\ldots i]:[i+1\ldots L]$ of the spins by
\begin{equation}
\ket{\Psi^{\mbox{\tiny [spin]}}} = \sum_{\mu_i} \lambda^{^{[i]}}_{\mu_i}~~ \ket{\Phi^{^{[1\ldots i]}}_{\mu_i}} \otimes \ket{\Phi^{^{[i+1 \ldots L]}}_{\mu_i}}.\label{eq:spinsd}
\end{equation}
Here $\lambda^{^{[i]}}$ is a diagonal matrix with non-negative diagonal entries $\lambda^{^{[i]}}_{\mu_i} \geq 0$, and $\{\ket{\Phi^{^{[1\ldots i]}}_{\mu_i}}\}$ and $\{\ket{\Phi^{^{[i+1 \ldots L]}}_{\mu_i}}\}$ are orthonormal bases for the two parts of the lattice.

\begin{figure}
  \includegraphics[width=5.5cm]{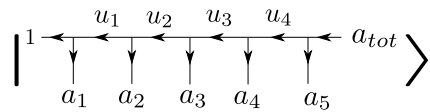}
\caption{Basis of a lattice system of $L=5$ anyons $\{a_1,a_2,a_3,a_4,a_5\}$ labelled by means of a \textit{fusion tree}. The latter corresponds to a sequence of pairwise fusing the anyons into a total anyon charge $a_\mathrm{tot}$. We have attached a redundant vacuum charge at the left of the fusion tree for convenience. The degrees of freedom of the lattice correspond to sets of the $L-1=4$ intermediate charges $\{u_1,u_2,u_3,u_4\}$ that are compatible with the fusion rules of the anyon model.\label{fig:intro2}}
\end{figure}

The generalization of the MPS formalism to lattice systems of anyons poses an interesting challenge. A system of anyons is not necessarily completely specified by giving the ``topological charge'' $a$ associated with each site of the lattice. For many species of anyons this information must be supplemented by the outcome of a number of non-local charge measurements, which can not be uniquely associated with single individual anyons. Thus, in contrast to a system of spins, a lattice system of $L$ anyons $\{a_1,a_2,\ldots a_L\}$ does not in general admit a description in terms of local Hilbert spaces associated with the lattice sites, and nor does the global Hilbert space admit a tensor product structure as in e.g.~\eref{eq:spinstate} and \eref{eq:spinsd}. Instead a basis is introduced by means of a \textit{fusion tree}, illustrated in \fref{fig:intro2} for a lattice of $L=5$ anyons $\{a_i\}_{i=1}^5$. A fusion tree specifies a sequence of pairwise fusions of the $L$ anyons (associated with the open edges of the tree) into a total anyon charge $a_\mathrm{tot}$ consistent with the \textit{fusion rules} of the anyon model. The fusion rules are a set of constraints on the charge outcomes arising from the fusion of an anyon pair. If the charges $\{a_i\}$ associated with the lattice sites are fixed, an anyonic lattice still possesses non-local degrees of freedom which
correspond to the set of possible charges $\{u_1,u_2,\ldots u_{L-1}\}$ that can appear on the internal edges of the fusion tree in arrangements consistent with the fusion rules.\cite{onlynonabel} The lack of a description in terms of local Hilbert spaces, and the subsequent necessity of a description in terms of a fusion tree, poses the key challenge in the direct simulation of anyonic systems. 

On the other hand, an anyonic lattice of this sort can be described by an enlarged Hilbert space that is spanned by the tensor product basis $\ket{u_1}\otimes\ket{u_2}\otimes\ldots\ket{u_{L-1}}$. This basis treats the intermediate charges $\{u_1,u_2,\dots u_{L-1}\}$ as effective spin degrees of freedom, but will in general contain unphysical states which are non-compliant with the fusion rules and whose existence must be suppressed. This mapping has been employed in DMRG \cite{Feiguin07} and Monte Carlo \cite{TB} studies of certain SU(2)$_k$ anyon models.
Direct simulation of anyon systems, without employing a mapping to spins, has also been performed using exact diagonalisation for up to 37 anyons. \cite{Feiguin07} More recently, the Multi-scale Entanglement Renormalization Ansatz (MERA) \cite{Vidal072,Vidal08} has been adapted \cite{Pfeifer10,Koenig10} to the fusion tree description of anyons by using \textit{charge-conserving} tensors that explicitly encode the fusion rules of the anyon model. This formalism offers a much broader avenue for simulation of anyonic lattice systems as it allows direct access to specific physical charge sectors in the Hilbert space for larger system sizes.

In this paper we describe how to adapt the MPS formalism to the fusion tree description of anyonic lattice systems. The anyonic MPS corresponds to a tensor network which is connected in the same way as the MPS for spin systems [\fref{fig:intro1}], but which is made of charge-conserving tensors.  
We also describe how to extend the TEBD algorithm to the anyonic MPS for the efficient simulation of time evolution under a local, anyonic charge-conserving Hamiltonian.

The key benefit of the anyonic MPS is that, by working directly in the fusion tree description of an anyonic system, it can be applied to study any anyon model given the description of that model in terms of the following parameters (see Appendix~\ref{sec:appA}):
\begin{enumerate}
\item The set of allowed anyon types or topological charges $a,b,c,\ldots$ 
\item The \emph{quantum dimension} $d_a$ associated with each charge $a$, analogous to the dimension of an irreducible representation in group theory. 
\item The fusion rules of the anyon model, encoded in the 3-index tensor $N_{ab}^c$, where $N^c_{ab}$ is the number of copies of charge $c$ appearing in the fusion product $a b$.
\item The three-index tensor $R^{ab}_c$ which describes the braiding of two anyons.
\item The 6-index tensor $(F^{abc}_d)_{ef}$, which relates different ways to fuse together three anyons via a relationship known as an $F$-move.
\end{enumerate}

In this paper, we will assume that all of these data are available for the anyon model of interest. For example, our method can be used to simulate anyons models with quantum symmetry SU(2)$_k$. In order to benchmark the anyonic TEBD algorithm we study first the ground state of an infinite chain of Ising anyons [described by SU(2)$_2$] and of Fibonacci anyons [described by SU(2)$_3$] subject to antiferromagnetic interaction. Both these models are critical and well studied. \cite{Feiguin07, TB, Pfeifer10,Koenig10} We then study the real time dynamics of an anyonic Hubbard-type model, \cite{ZLSPB,LZBPW,Lehman2013} and demonstrate that the transport behaviour depends on the presence or absence of topological disorder which our method can accommodate in a straightforward manner.

The rest of the paper is organized as follows. Section~\ref{sec:one} introduces the Schmidt decomposition and matrix product decomposition of pure anyonic states and outlines the generalization of TEBD algorithm to the anyonic MPS.
Section~\ref{sec:three} contains the numerical results. Appendix~\ref{sec:appAA} recapitulates the derivation of the standard Schmidt decomposition for spin systems. The derivation of the anyonic Schmidt decomposition presented in Sec.~\ref{ssec:bidecom} follows the same sequence of steps described in Appendix~\ref{sec:appAA} but adapts each step to the anyonic setting. The basic terminology and graphical notation pertaining to anyon models as used in this paper is summarised in Appendix~\ref{sec:appA}. Appendix~\ref{sec:atebd} describes the step-by-step implementation of the anyonic TEBD algorithm. Appendix~\ref{sec:general} describes some straightforward generalizations of the anyonic MPS formalism presented in this paper, and also its specialization to study systems with conserved symmetry charges.

\subsection{Notation convention and assumptions}

In this paper, we essentially follow the graphical notation for anyon models described in Refs.~\onlinecite{BondersonThesis} and~\onlinecite{kitaev2006}. However, we find it convenient to rotate the graphical representation of fusion trees in Ref.~\onlinecite{BondersonThesis} counterclockwise to mimic the graphical representation of the MPS [\fref{fig:intro1}].
Exploiting the fact that for any anyonic model there necessarily exists a vacuum charge with trivial fusion rules, we sometimes attach this charge to the left of the fusion tree for convenience, as in \fref{fig:intro2}. When we do so for a fusion tree of $L$ anyons the number of intermediate charges (those appearing on the internal edges) in the fusion tree is $L-1$, whereas there would be only $L-2$ intermediate charges if we did not introduce the trivial vacuum charge.

For the purpose of clearer demonstration, we have made certain simplifying assumptions in this paper. First, we assume that the total fusion charge [$a_\mathrm{tot}$ in \fref{fig:intro2}] assumes only one value (this condition arises naturally when describing a pure state).
Second, we assume that each site of the anyonic lattice is described purely by a charge label from the anyon model, with no additional degeneracies or auxiliary degrees of freedom.
Third, we restrict to \textit{multiplicity-free} anyon models where the components $N_{ab}^c$ [\eref{eq:fusionrules} in Appendix~\ref{sec:appA}] only take values 0 or 1. Appendix~\ref{sec:general} describes how the formalism can be generalized in a straightforward way to relax the latter two of these three assumptions.

Finally, although it is common practice to consider only anyonic states with $a_\mathrm{tot}=1$  [\fref{fig:intro2}], we do not assume any particular value of $a_\mathrm{tot}$ throughout the paper except in the construction of the anyonic Schimdt decomposition presented in Sec.~\ref{ssec:bidecom} (where we set $a_\mathrm{tot}=1$ for convenience) and in Sec.~\ref{sec:three} which presents the numerical results. Our methodology may therefore readily be applied to systems with non-trivial total charge.

\section{Anyonic matrix product states\label{sec:one}}
Consider a one dimensional lattice $\mathcal{L}$ made of $L$ sites that are fixed on a line and populated by anyons $\{a_1,a_2,\ldots,a_L\}$. Denote by $\veclat{L}$ the Hilbert space that describes lattice $\mathcal{L}$. A basis is introduced in $\veclat{L}$ by means of a fusion tree (illustrated in \fref{fig:intro2}). We may also denote a fusion tree basis by explicitly listing the sequence of fusions in the tree; for example, we may denote the basis depicted in \fref{fig:intro2} as 
\begin{equation*}
\ket{\fuse{1}{a_1}{u_1},\fuse{u_1}{a_2}{u_2},\ldots,\fuse{u_4}{a_5}{a_\mathrm{tot}}}.
\end{equation*}

In this paper we are interested in states $\ket{\Psi}$ that have a well defined total charge $a_\mathrm{tot}$. State $\ket{\Psi}$ can be expanded as
\begin{equation}
\ket{\Psi} =\!\! \sum  \Psi_{\{a_i\}\{u_j\}} \ket{\fuse{1}{a_1}{u_1},\fuse{u_1}{a_2}{u_2},\ldots,\fuse{u_{L-1}}{a_L}{a_\mathrm{tot}}},\label{eq:genstate}
\end{equation}
where $\Psi_{\{a_i\}\{u_j\}}$ are complex coefficients and the sum is over all sets of \textit{compatible} charges $\{a_i\}_{i=1}^L$ and $\{u_j\}_{j=1}^{L-1}$, namely, sets of charges resulting in a valid fusion tree.

The space $\veclat{L}$ decomposes as
\begin{equation}
\veclat{L} \cong \bigoplus_{a_\mathrm{tot}} \mathbb{V}^{(\mathcal{L})}_{a_\mathrm{tot}},\label{eq:totspace}
\end{equation}
where $\mathbb{V}^{(\mathcal{L})}_{a_\mathrm{tot}}$ is a subspace of states in $\veclat{L}$ that have a well defined total charge $a_\mathrm{tot}$. The dimension of subspace $\mathbb{V}^{(\mathcal{L})}_{a_\mathrm{tot}}$ is equal to the number of ways $n_{a_\mathrm{tot}}$ in which the total charge $a_\mathrm{tot}$ can be obtained by fusing together the $L$ anyons. We also refer to $n_{a_\mathrm{tot}}$ as the \textit{degeneracy} of total charge $a_\mathrm{tot}$ in the decomposition of \eref{eq:totspace}. 

State $\ket{\Psi}$ can be expanded as
\begin{equation}
\ket{\Psi} = \sum_{a_\mathrm{tot}}\sum_{t_{a_\mathrm{tot}}} \Psi_{a_\mathrm{tot} t_{a_\mathrm{tot}}} \ket{a_\mathrm{tot} t_{a_\mathrm{tot}}},\label{eq:sing}
\end{equation}
in accordance with the decomposition (\ref{eq:totspace}). Here $\ket{a_\mathrm{tot} t_{a_\mathrm{tot}}}$ denotes an orthonormal basis in the subspace $\mathbb{V}^{(\mathcal{L})}_{a_\mathrm{tot}}$ such that $ t_{a_\mathrm{tot}} = 1,2,\ldots n_{a_\mathrm{tot}}$, the \textit{degeneracy index} of charge $a_\mathrm{tot}$, enumerates the different labellings of the fusion tree that are compatible with the given value of charge $a_\mathrm{tot}$.

\subsection{Anyonic bipartite decomposition\label{ssec:bidecom}}
In this subsection we describe generic bipartite decompositions and the Schmidt decomposition of an anyonic state $\ket{\Psi}$. The latter plays an instumental role in constructing the matrix product decomposition of $\ket{\Psi}$. The reader interested mostly in the definition of the anyonic MPS and in the implementation of the anyonic TEBD algorithm may skip the following technical discussion and proceed directly to Sec.~\ref{ssec:amps}. 

Consider a bipartition of $\mathcal{L}$ into sublattices $\mathcal{A}$ and $\mathcal{B}$ that consist of anyons $\{a_1,a_2,\ldots,a_i\}$ and $\{a_{i+1},a_{i+2},\ldots,a_L\}$ respectively. Denote by $\mathbb{V}^{(\mathcal{A})}$ and $\mathbb{V}^{(\mathcal{B})}$ the vector spaces that describe $\mathcal{A}$ and $\mathcal{B}$ respectively. We have
\begin{equation}
\mathbb{V}^{(\mathcal{A})} \cong \bigoplus_a \mathbb{V}^{(\mathcal{A})}_a,~~~\mathbb{V}^{(\mathcal{B})} \cong \bigoplus_b \mathbb{V}^{(\mathcal{B})}_b,\label{eq:ABdecom}
\end{equation}
where $\mathbb{V}^{(\mathcal{A})}_a$ and $\mathbb{V}^{(\mathcal{B})}_b$ are the degeneracy spaces [\eref{eq:totspace}] of total charges $a$ and $b$ for sublattices $\mathcal{A}$ and $\mathcal{B}$ respectively. The total space $\mathbb{V}^{(\mathcal{L})}$ is the tensor product of $\mathbb{V}^{(\mathcal{A})}$ and $\mathbb{V}^{(\mathcal{B})}$, and decomposes according to
\begin{equation}
\mathbb{V}^{(\mathcal{L})}_{a_\mathrm{tot}}  \cong \bigoplus_{a,b : N_{ab}^{a_\mathrm{tot}}=1} \left(\mathbb{V}^{(\mathcal{A})}_a \otimes \mathbb{V}^{(\mathcal{B})}_b\right),\label{eq:bidecom}
\end{equation}
where $\mathbb{V}^{(\mathcal{L})}_{a_\mathrm{tot}}$ is the degeneracy space of total charge $a_\mathrm{tot}$  in \eref{eq:totspace} and the direct sum is over charges $a$ and $b$ that are compatible with $a_\mathrm{tot}$ according to the fusion rules, that is, the set of charges $\{a,b\}$ which satisfy $N_{ab}^{a_\mathrm{tot}}=1$. Here $N$ is the 3-index tensor that encodes the fusion rules of the anyon model, defined according to \eref{eq:fusionrules} in Appendix~\ref{sec:appA}. Note that, in general, there may be several $(a,b)$ that contribute to the degeneracy of total charge $a_\mathrm{tot}$ in the decomposition \eeref{eq:bidecom}.
\begin{figure}
\includegraphics[width=8cm]{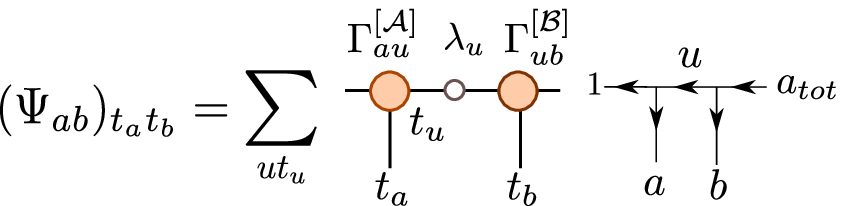}
\caption{Canonical bipartite decomposition of an anyonic state $\ket{\Psi}$ with definite charge $a_\mathrm{tot}$. The coefficients $(\Psi_{ab})_{t_at_b}$ in \eref{eq:bistate0} are encoded in \textit{degeneracy tensors} $\Gamma^{^{[\mathcal{A}]}}_{au}, \lambda_u$ and $\Gamma^{^{[\mathcal{B}]}}_{ub}$ with components  $(\Gamma^{^{[\mathcal{A}]}}_{au})_{t_at_u}, (\lambda_u)_{t_u}$, and $(\Gamma^{^{[\mathcal{B}]}}_{ub})_{t_ut_b}$ respectively by means of the singular value decomposition \eeref{eq:blocksvd}. Charges $a,b$ and $u$ fulfill the fusion rules (\ref{eq:aub}) that are depicted here by the fusion tree. \label{fig:bipartite}}
\end{figure}

Notice how the introduction of degeneracy spaces allows for a decomposition of the anyonic Hilbert space $\mathbb{V}^{(\mathcal{L})}$ as a direct sum of \textit{tensor product} spaces [\eref{eq:bidecom}]. In the remainder of this section, we describe how this decomposition can be exploited to construct a bipartite decomposition and the matrix product decomposition of the anyonic state $\ket{\Psi}$ in Eqs.~(\ref{eq:genstate}) and (\ref{eq:sing}).

Let $\ket{at_a}$ and $\ket{bt_b}$ denote an orthonormal basis in $\mathcal{A}$ and $\mathcal{B}$ respectively. Then in accordance with the decomposition (\ref{eq:bidecom}), we can choose a basis $\ket{a_\mathrm{tot} t_{a_\mathrm{tot}}}$ [\eref{eq:sing}] in the total space $\mathbb{V}^{(\mathcal{L})}$ that factorizes as
\begin{equation}
\ket{a_\mathrm{tot} t_{a_\mathrm{tot}}} = N_{ab}^{a_\mathrm{tot}}\ket{at_a}\otimes \ket{bt_b},\label{eq:factorbasis}
\end{equation}
where the fusion rule $a,b \rightarrow a_\mathrm{tot}$ enforces a total charge $a_\mathrm{tot}$. A generic bipartite decomposition of state $\ket{\Psi}$ according to bipartition $\mathcal{A}:\mathcal{B}$ reads as
\begin{align}
\ket{\Psi} &= \sum_{at_a} \sum_{bt_b}\Psi_{at_a, bt_b} N_{ab}^{a_\mathrm{tot}}\ket{at_a}\otimes \ket{bt_b}. \label{eq:bistate0}
\end{align}

Next, we introduce the Schmidt decomposition of $\ket{\Psi}$ according to the bipartition $\mathcal{A}:\mathcal{B}$. Our derivation of the anyonic Schmidt decomposition follows the same sequence of steps involved in the standard derivation of the Schmidt decomposition for spins systems, which is recapitulated in Appendix~\ref{sec:appAA}. We refer the reader to Appendix~\ref{sec:appAA} as an aid to understanding the following derivation.

Without loss of generality, \cite{CBD} we now specialize to trivial total charge ($a_\mathrm{tot}=1$) for simplicity. Components $\Psi_{at_a, bt_b}$ in \eref{eq:bistate0} can be organized as a matrix $\Psi$ where the paired indices $a,t_a$ and $b,t_b$ label the rows and columns respectively. Since $a_\mathrm{tot}=1$, the fusion rules $N_{ab}^{a_\mathrm{tot}}=1$ imply that charge $b$ is the \textit{dual} of charge $a$ (denoted as $b=a^*$) and therefore matrix $\Psi$ is block diagonal as
\begin{equation}
\Psi = \!\!\!\!\!\!\!\!\bigoplus_{a,b:N_{ab}^{1}=1}\!\!\!\! \Psi_{ab},\label{eq:biblocks}
\end{equation}
where $\Psi_{ab}$ is a matrix block with components $(\Psi_{ab})_{t_at_b}$. Notice that we now denote the components of $\Psi$ as $(\Psi_{ab})_{t_at_b}$ instead of $\Psi_{at_a, bt_b}$ to explicitly indicate the block structure (\ref{eq:biblocks}).
\begin{figure}
\includegraphics[width=6cm]{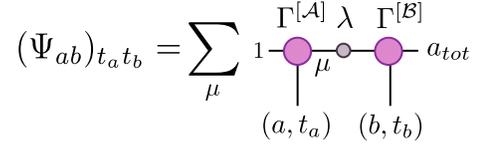}
\caption{A compact graphical representation of the canonical bipartite decomposition of \fref{fig:bipartite} in terms of \textit{charge-conserving} tensors $\Gamma^{^{[\mathcal{A}]}}$, $\lambda$, and $\Gamma^{^{[\mathcal{B}]}}$, which decompose in terms of degeneracy tensors $\Gamma^{^{[\mathcal{A}]}}_{au}$, $\lambda_u$, and $\Gamma^{^{[\mathcal{B}]}}_{ub}$ respectively according to \eref{eq:allblock}. 
\label{fig:bipartite1}}
\end{figure}

Consider the singular value decomposition of block $\Psi_{ab}$ of \eref{eq:biblocks},
\begin{equation}
\Psi_{ab} = \Gamma^{^{[\mathcal{A}]}}_{au} \lambda_u \Gamma^{^{[\mathcal{B}]}}_{ub},\label{eq:blocksvd}
\end{equation}
or in terms of components (see \fref{fig:bipartite})
\begin{equation}
(\Psi_{ab})_{t_at_b} = \sum_{t_u}(\Gamma^{^{[\mathcal{A}]}}_{au})_{t_at_u} (\lambda_u)_{t_u} (\Gamma^{^{[\mathcal{B}]}}_{ub})_{t_ut_b}.\label{eq:blocksvd1}
\end{equation}
where $\Gamma^{^{[\mathcal{A}]}}_{au}$ and $\Gamma^{^{[\mathcal{B}]}}_{ub}$ are unitary matrices,
\begin{align}
\Gamma^{^{[\mathcal{A}]}}_{au} (\Gamma^{^{[\mathcal{A}]}}_{au})^\dagger = \Gamma^{^{[\mathcal{B}]}}_{ub}({\Gamma^{^{[\mathcal{B}]}}_{ub}})^\dagger  = \mathbb{I}.
\end{align}
and $\lambda_u$ is a diagonal matrix with non-negative diagonal entries, $(\lambda_u)_{t_u} \geq 0$. The charges $a,u$ and $b$ satisfy the fusion rules: 
\begin{equation}
N_{1a}^u=1,~~~N_{ub}^1=1.\label{eq:aub}
\end{equation}
The fusion rule $N_{1a}^u=1$ simply implies that $a=u$, but we have introduced a new label $u$ to clearly distinguish the corresponding degeneracy indices $t_a$ and $t_u$, which are independent of each other in the following discussion. The fusion rule $N_{ub}^1=1$ implies that $b=u^*$. Therefore, charges $a$ and $b$ can be uniquely determined if $u$ is specified.

The SVD of the total matrix $\Psi$ is (see \fref{fig:bipartite1})
\begin{equation}
\Psi = \Gamma^{^{[\mathcal{A}]}} \lambda \Gamma^{^{[\mathcal{B}]}},
\end{equation}
where matrices $\Gamma^{^{[\mathcal{A}]}}, \lambda$ and $\Gamma^{^{[\mathcal{B}]}}$ are block diagonal,
\begin{equation}
\Gamma^{^{[\mathcal{A}]}} = \!\!\!\!\!\!\bigoplus_{au:N_{1a}^u=1} \!\!\!\!\Gamma^{^{[\mathcal{A}]}}_{au},~~~\lambda = \bigoplus_u \lambda_u,~~~ \Gamma^{^{[\mathcal{B}]}} = \!\!\!\!\!\!\bigoplus_{ub:N_{ub}^1=1} \!\!\!\!\Gamma^{^{[\mathcal{B}]}}_{ub},\label{eq:allblock}
\end{equation}
with the blocks $\Gamma^{^{[\mathcal{A}]}}_{au}, \lambda_u$ and $\Gamma^{^{[\mathcal{B}]}}_{ub}$ obtained according to \eref{eq:blocksvd}. We also say that matrices $\Gamma^{^{[\mathcal{A}]}}, \lambda$ and $\Gamma^{^{[\mathcal{B}]}}$ which have a block structure compatible with the fusion rules are \textit{charge-conserving}, meaning that they transform a state with given anyonic charge to a state with the same charge.

Using \eref{eq:blocksvd1} in \eref{eq:bistate0} and summing over $t_a$ and $t_b$ we obtain
\begin{equation}
\ket{\Psi} = \sum_{u}\sum_{t_u} (\lambda_{u})_{t_u} \ket{\Phi^{^{[\mathcal{A}]}}_{ut_u}}\otimes \ket{\Phi^{^{[\mathcal{B}]}}_{u^*t_{u^*}}}, \label{eq:schmidt0}
\end{equation}
where we have replaced $a=u$ and $b=u^*$ using (\ref{eq:aub}) and defined the orthonormal vectors
\begin{equation}
\begin{split}
\ket{\Phi^{^{[\mathcal{A}]}}_{ut_u}} = \sum_{t_a} (\Gamma^{^{[\mathcal{A}]}}_{au})_{t_at_u} N_{1a}^u\ket{at_a},\\
\ket{\Phi^{^{[\mathcal{B}]}}_{u^*t_{u^*}}} = \sum_{t_b} (\Gamma^{^{[\mathcal{B}]}}_{ub})_{t_ut_b} N_{ub}^1\ket{bt_b}.
\end{split}
\end{equation}
We write \eref{eq:schmidt0} more succintly by introducing the paired index $\mu = (u,t_u)$ and its dual $\mu^* = (u^*,t_{u^*})$,
\begin{align}
\ket{\Psi} &= \sum_{\mu} \lambda_{\mu} \ket{\Phi^{^{[\mathcal{A}]}}_\mu}\otimes \ket{\Phi^{^{[\mathcal{B}]}}_{\mu^*}}\label{eq:aschmidt}
\end{align}
with concise graphical representation given in \fref{fig:bipartite1}. Equation~(\ref{eq:aschmidt}) is the anyonic Schmidt decomposition. The similarity between \eref{eq:aschmidt} and \eref{eq:spinsd} is apparent, however, note the distinction: here index $\mu$ is a charge-degeneracy pair [in accordance with (\ref{eq:ABdecom})], and the Schmidt bases in $\mathcal{A}$ and $\mathcal{B}$ are labeled by $\mu$ and its dual $\mu^*$ respectively to constrain the total charge of the bipartite anyonic state $\ket{\Psi}$ to $a_\mathrm{tot}=1$.

The norm of state $\ket{\Psi}$ is given as
\begin{equation}
\braket{\Psi}{\Psi} = \sum_{u}d_u \left[\sum_{t_u} (\lambda_u)^{^2}_{t_u}\right].\label{eq:normm}
\end{equation}
The anyonic Schmidt decomposition is a useful tool to probe bipartite entanglement in an anyonic state $\ket{\Psi}$.\cite{hikami2008,pfeifer2013} In analogy with spin or bosonic systems (see Appendix~\ref{sec:appAA}), we define the Von Neumann entanglement entropy $S(\mathcal{A}:\mathcal{B})$ of parts $\mathcal{A}$ and $\mathcal{B}$ of a pure anyonic state $\ket{\Psi}$ as
\begin{equation}
S(\mathcal{A}:\mathcal{B}) = -\sum_{u}d_u\left[\sum_{t_u} (\lambda_{u})^{^2}_{t_u} \mbox{log}(\lambda_{u})^{^2}_{t_u}\right],
\end{equation}
where $d_u$ is the quantum dimension of charge $u$.

\subsection{Anyonic matrix product decomposition}\label{ssec:amps}
\begin{figure}
\includegraphics[width=\columnwidth]{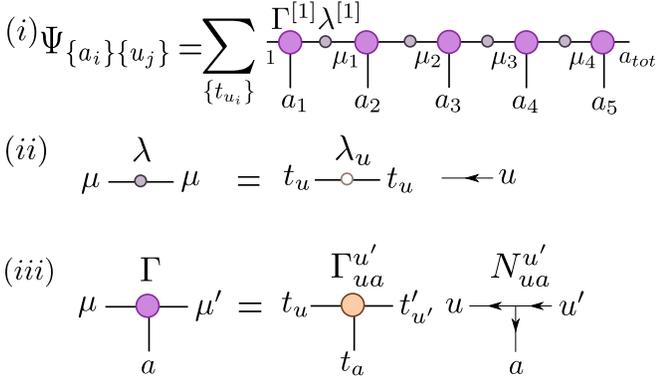}
\caption{$(i)$ Matrix product decomposition of state $\ket{\Psi}$ of 5 anyons. The coefficients $\Psi_{\{a_i\}\{u_j\}}$ in \eref{eq:genstate} are encoded in anyonic charge-conserving tensors $\Gamma^{^{[i]}}$ and $\lambda^{^{[i]}}$. $(ii)$ For a fixed charge $u$, each diagonal matrix $\lambda$ decomposes into diagonal degeneracy matrix $\lambda_{u}$ and the fusion tensor $N_{u1}^{u}$ (we have suppressed the vacuum label in the figure). $(iii)$ For fixed charges $a,u$ and $u'$, each tensor $\Gamma$ decomposes into a degeneracy tensor $\Gamma_{ua}^{u'}$ and a fusion tensor $N_{ua}^{u'}$. See \eref{eq:gamma}. Note that the canonical bipartite decomposition of \fref{fig:bipartite1} may be regarded as a 2-site matrix product decomposition of $\ket{\Psi}$. \label{fig:anyonicmps}}
\end{figure}

More generally, components $\Psi_{\{a_i\}\{u_j\}}$ in \eref{eq:genstate} can be encoded as a \textit{Matrix Product State} [see \fref{fig:anyonicmps}(i)]; that is,
\begin{equation}
\Psi_{\{a_i\}\{u_j\}} = \sum_{\{t_{u_j}\}} \Gamma^{^{[1]}\mu_1}_{1 a_1} \lambda^{^{[1]}}_{\mu_1} \Gamma^{^{[2]}\mu_2}_{\mu_1 a_2} \lambda^{^{[2]}}_{\mu_2} \ldots \lambda^{^{[L-1]}}_{\mu_{L-1}} \Gamma^{^{[L]}a_\mathrm{tot}}_{\mu_{L-1} a_L}.\label{eq:amps}
\end{equation}
Here index $\mu_i$ is a charge-degeneracy pair $\mu_i = (u_i,t_{u_i})$, $\lambda^{^{[i]}}$ is the diagonal matrix that appears in the anyonic Schmidt decompositon according to the bipartition $[1\ldots i]:[i+1\ldots L]$ and tensors $\Gamma^{^{[i]}}$ relate the Schmidt basis for consecutive bipartitions as
\begin{align}
&\ket{(\Phi^{^{[i\ldots L]}}_{u})_{t_u}}
 = (\Gamma_{ua_i}^{^{[i]}u'})_{t_ut_{u'}} (\lambda^{^{[i]}}_{u'})_{t_{u'}}N_{ua_i}^{u'}~~\ket{(\Phi^{^{[i+1\ldots L]}}_{u'})_{t_{u'}}}.
\end{align}

Analogous to \eref{eq:allblock}, tensors $\lambda$ and $\Gamma$ can be decomposed in accordance with the fusion rules as [see \fref{fig:anyonicmps}(ii)-(iii)]
\begin{equation}
\lambda = \bigoplus_u \lambda_u,~~~~~\Gamma = \!\!\!\!\!\bigoplus_{auu':N_{ua}^{u'}=1}\!\!\!\! \Gamma_{ua}^{u'},\label{eq:gamma}
\end{equation}
where $\lambda_u$ is a diagonal \textit{degeneracy matrix} with diagonal entries $(\lambda_u)_{t_u} \geq 0$ and $\Gamma_{ua}^{u'}$ is a \textit{degeneracy tensor} with components $(\Gamma_{ua}^{u'})_{t_at_u}^{t_{u'}}$. The decompositions (\ref{eq:gamma}) imply that $\lambda$ and $\Gamma$ correspond to linear maps that conserve anyonic charge. 

We refer to the decomposition \eref{eq:amps} in terms of charge-conserving tensors as \textit{the anyonic MPS}. By working with a fusion tree that mimics the tensor network structure of the MPS, manifesting as the visual similarity between \fref{fig:anyonicmps} and \fref{fig:intro2}, we obtain a direct correspondence between the fusion tree description and the MPS description of an anyonic state. Namely, the physical indices and the bond indices of the MPS are labelled by charges $\{a_i\}_{i=1}^L$ and $\{u_i\}_{i=1}^{L-1}$ that appear on the open and internal edges of the fusion tree respectively. This means that for given charges $\{a_i\},\{u_j\}$ the coefficients $\Psi_{\{a_i\}\{u_j\}}$ in \eref{eq:genstate} can be recovered from the anyonic MPS by fixing these charges on the respective MPS indices, decomposing each charge-conserving tensor according to (\ref{eq:gamma}) and multiplying together the degeneracy tensors.

Next, we explain how the TEBD algorithm is adapted to the anyonic MPS by ensuring that the fusion constraints encoded in the MPS tensors are preserved during time evolution.

\subsection{Simulation of time evolution\label{sec:two}}
In this section we describe how to simulate the time evolution of an anyonic matrix product state $\ket{\Psi(0)}$,
\begin{equation}
\ket{\Psi(t)} = U(t) \ket{\Psi(0)},~~~U(t) = e^{-iH t}\label{eq:timeevolve}
\end{equation}
where $H:\mathbb{V}^{(\mathcal{L})} \rightarrow \mathbb{V}^{(\mathcal{L})}$ is a \textit{local} and \textit{charge-conserving} Hamiltonian. Here \textit{local} implies that $H$ is a sum of finite range interactions, for example,
\begin{equation}
H = \sum_{i} h^{^{[i, i+1]}},\label{eq:localham}
\end{equation}
and \textit{charge-conserving} implies that each nearest neighbour term $h^{^{[i, i+1]}}$ is block diagonal in the fusion space $\mathbb{V}^{^{[i,i+1]}}$ of anyons $a_i$ and $a_{i+1}$, that is,
\begin{align}
\mathbb{V}^{^{[i,i+1]}} = \!\!\!\!\!\!\!\!\!\!\!\bigoplus_{p:N_{a_i,a_{i+1}}^p=1} \!\!\!\!\!\!\!\!\!\!\mathbb{V}^{^{[i,i+1]}}_p,~~~~~
h^{^{[i, i+1]}} = \bigoplus_p~h^{^{[i,i+1]}}_{p},\label{eq:localham1}
\end{align}
where $p$ is the charge obtained by fusing $a_i$ and $a_{i+1}$.

\begin{figure}
\includegraphics[width=\columnwidth]{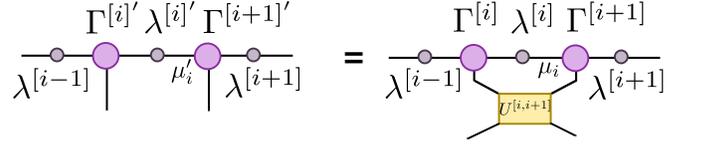}
\caption{Main step of the TEBD algorithm. Tensors $\Gamma^{^{[i]}}, \lambda^{^{[i]}}$ and $\Gamma^{^{[i+1]}}$ are locally updated after absorbing the charge-conserving time-evolution gate $U^{^{[i,i+1]}}$. \label{fig:mainstep}} 
\end{figure}

Following Ref.~\onlinecite{Vidal03} we perform a Trotter decomposition of $U(t)$ in \eref{eq:timeevolve} over a sequence of small time steps $\delta t = t/n,~n\gg1$,
\begin{equation}
U(t) \approx \left[\prod_{i\in~even} U^{^{[i,i+1]}}(\delta t) \prod_{i\in~odd} U^{^{[i,i+1]}}(\delta t)\right]^n + O(\delta t^2).
\end{equation}
Each 2-site gate $U^{^{[i,i+1]}}(\delta t)$ decomposes as per \eref{eq:localham1}:
\begin{equation}
U^{^{[i,i+1]}}(\delta t)= \bigoplus_p U^{^{[i,i+1]}}_p.
\end{equation}
The main step of the (anyonic) TEBD algorithm is to update the MPS after applying a 2-site gate $U^{^{[i,i+1]}}(\delta t)$, as depicted formally in \fref{fig:mainstep}. As explained in Ref.~\onlinecite{Vidal03} for a non-anyonic MPS, this update comprises of certain tensor contractions and a matrix singular value decomposition. For the anyonic MPS the goal is to ensure that the updated tensors ${\Gamma^{^{[i]}}}', {\lambda^{^{[i]}}}'$ and ${\Gamma^{^{[i+1]}}}'$ are charge-conserving, having a block structure that is compatible with the fusion rules. This is achieved by decomposing the anyonic MPS tensors into degeneracy and fusion parts according to \eref{eq:gamma}. The step by step details of how to enact the update of \fref{fig:mainstep} for the anyonic MPS is explained in Appendix~\ref{sec:atebd}.

In practical simulations, a truncation is made after the singular value decomposition step of the update by retaining only a fixed number $\chi$ of singular values ${\lambda^{^{[i]}}}'$. During this truncation the index $\mu_i$ is replaced by an index $\mu'_i$ where the
degeneracies $t'_{u'_i}$ of the charges $u'_i$ in $\mu'_i$ are chosen such that the norm \eeref{eq:normm} of the updated MPS $\ket{\Psi}$ is maximized, subject to the limitation imposed by the value of $\chi$:
\begin{equation}
\braket{\Psi}{\Psi} = \sum_{u'_i}d_{u_i'}\left[\sum_{t_{u'_i}} ({\lambda^{^{[i]}}_{u_i'}}')^{^2}_{t_{u_i'}}\right],~~~ \sum_{u_i'} |t_{u_i'}| = \chi.\label{eq:chi00}
\end{equation}
The degeneracy of a given charge in $\mu'_i$ need not therefore coincide with the degeneracy of the equivalent charge in $\mu_i$.

Let us denote by $n_{max} = \mbox{max}\{|t_{u'_i}|\}_{u'_i}$ the maximum degeneracy associated with any charge $u'_i$ that appears on the bond indices of the anyonic MPS. When the amount of entanglement in the ground state $\ket{\Psi}$ of $H$ is limited, namely, when $n_{max}$ is bounded and does not scale with system size $L$, the anyonic MPS allows for an extremely efficient description of $\ket{\Psi}$ in terms of approximately $O(L n_{max}^2)$ coefficients. The maximum degeneracy $n_{max}$ also controls the computational CPU cost incurred by the anyonic TEBD algorithm: the CPU cost scales approximately as $O(n^3_{max})$, being dominated by the cost of the singular value decomposition step of the algorithm.
\begin{figure}
  \includegraphics[width=7.5cm]{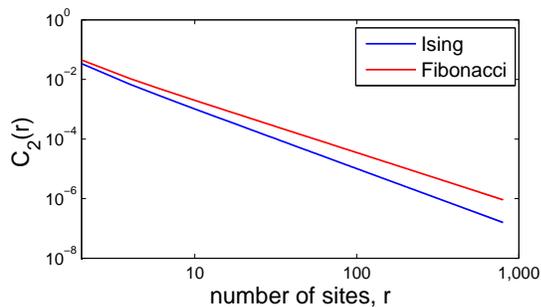}
\caption{Two point correlator $C_2(r)$, \eref{eq:2point}, between sites $i$ and $i+r$ ($r=2\ell~,\ell=1,2,\ldots$) of the energy density $h^{[i,i+1]}$ of the ground state of the infinite antiferromagnetic chain of Ising anyons and Fibonacci anyons, approximated by means of an anyonic MPS with $\chi=200$.\label{fig:correlators}}
\end{figure}

\section{Benchmark results\label{sec:three}}

To demonstrate the effectiveness of the algorithm, we applied it to the study of two different types of interacting (quasi-) one-dimensional models of interacting anyons.
\subsection{Infinite chain of anyons with antiferromagnetic interactions\label{ssec:heisenberg}}
We considered an infinite chain of anyons $a_{i}$ with a nearest neighbour antiferromagnetic interaction. That is, for the nearest neighbour fusion process
\begin{equation}
a_i \times a_{i+1} \longrightarrow p, 
\end{equation}
the Hamiltonian favours fusion to the vacuum, $p=1$. We studied two different anyon models: Ising anyons and Fibonacci anyons (see Sec.~\ref{ssec:basicdata}). For the Ising anyon model, a $\sigma$ anyon is placed at each site $i$, that is,~$a_i= \sigma$. Two neighbouring anyons $a_i= \sigma$ and $a_{i+1} = \sigma$ may fuse to the vacuum $p=1$ or to $p=\psi$. The 2-site Hamiltonian $h^{[i,i+1]}$ \eeref{eq:localham} is given by two $1 \times 1$ matrices acting on the two sectors of the fusion space,
\begin{equation}
h^{^{[i,i+1]}}_{p=1} = (-1), ~~~~~h^{^{[i,i+1]}}_{p=\psi} = (0).\label{eq:ising}
\end{equation}
Similarly, for the Fibonacci anyon model a $\tau$ anyon is placed at each site. Two neighbouring $\tau$ anyons may fuse either to the vacuum $p=1$ or to $p=\tau$, and the 2-site Hamiltonian is given by
\begin{equation}
h^{^{[i,i+1]}}_{p=1} = (-1), ~~~~~h^{^{[i,i+1]}}_{p=\tau} = (0).\label{eq:fib}
\end{equation}
The Hamiltonians (\ref{eq:ising}--\ref{eq:fib}) can be mapped onto spin-$1/2$ XXZ chains with a quantum group $SU(2)_k$ symmetry with $k=2$ for the Ising model and $k=3$ for the Fibonacci model,\cite{IsingSU2} and this symmetry is made manifest in the XXZ model by the addition of non-Hermitian terms on the boundaries. \cite{Gomez} In the thermodynamic limit the systems are described by $(k-1)$-th minimal models of conformal field theory (CFT), and the ground states of the Hamiltonians (\ref{eq:ising}--\ref{eq:fib}) are described by the Ising CFT (with central charge equal to $1/2$) and Tricritical Ising CFT (with central charge equal to $7/10$) respectively. Both models have been studied previously using DMRG \cite{Feiguin07,Trebst08}
and valence bond Monte Carlo. \cite{TB} The Fibonacci model (\ref{eq:fib}) has also been studied using the anyonic MERA. \cite{Pfeifer10,Koenig10}
\begin{figure}
  \includegraphics[width=7.5cm]{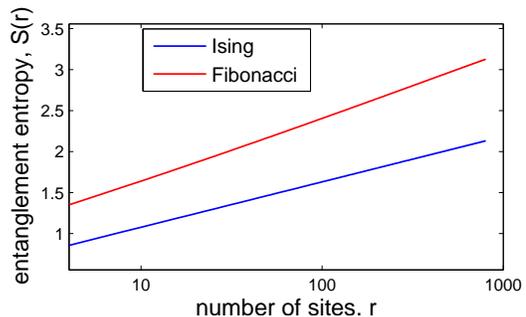}
\caption{Scaling of entanglement entropy $S(r)$, \eref{eq:rho}, of a block of $r=2\ell~(\ell=1,2,\ldots)$ sites for the ground state of the infinite antiferromagnetic chain of Ising anyons and Fibonacci anyons, approximated by means of an anyonic MPS with $\chi=200$.\label{fig:entropy}}
\end{figure}

\begin{table}
\centering 
\begin{tabular}{c| c| c} 
\multicolumn{3}{ c }{Ising anyons} \\
\hline\hline 
& \textit{charges} & \textit{degeneracy} \\ [0.5ex] 
\hline
\multirow{2}{*}{even}   & 1 & 100 \\
 & $\psi$ & 100  \\ \hline
odd & $\sigma$ & 200 \\ 
\hline 
\end{tabular}
~~~
\begin{tabular}{c| c| c} 
\multicolumn{3}{ c }{Fibonacci anyons} \\
\hline\hline 
& \textit{charges} & \textit{degeneracy} \\ [0.5ex] 
\hline
\multirow{2}{*}{even}   & 1 & 76 \\
 & $\tau$ & 124  \\ \hline
\multirow{2}{*}{odd}   & 1 & 76 \\
 & $\tau$ & 124  \\
\hline 
\end{tabular} 
\caption{Charges $c$ with degeneracies $|t_{c}|$ that contribute to the even and odd bipartitions \cite{evenodd} of the ground state approximated with the constraint $\chi=200$ in \eref{eq:chi}. \label{table:charges}}
\end{table}

We used the anyonic TEBD algorithm to approximate the ground state $\ket{\Psi_{gs}}$ of the two models by means of imaginary time evolution,
\begin{equation}
\ket{\Psi_{gs}} = \lim_{t \to \infty} e^{-Ht} \ket{\Psi(0)},
\end{equation}
and imposed the constraint $\chi \leq 200$ in \eref{eq:chi00}. Table~\ref{table:charges} lists the charges $c$ with degeneracies $|t_{c}|$ that contribute to the even and odd bipartitions of the resulting state with the constraint $\chi=200$. \cite{evenodd}

\begin{table}
\centering 
\begin{tabular}{c| c| c} 
\hline\hline 
bond dimension ($\chi$) & \mbox{Ising anyons} & Fibonacci anyons \\ [0.5ex] 
\hline
50 & -0.81830988[4] & -0.76393[1] \\
200 & -0.818309886[0] & -0.76393202[1]  \\
$\infty$ (exact) \cite{TB} & -0.81830988618 & -0.7639320225 \\ 
[1ex]
\hline 
\end{tabular}
\caption{
Energy per site of an infinite antiferromagnetic chain of Ising anyons and Fibonacci anyons obtained using an anyonic MPS with bond dimensions $\chi=50,200$, [.] indicates the first significant digit that differs from the exact energy. The exact energies are listed only upto 10 significant digits.\label{table:gsenergy} 
}
\end{table}
We obtained an accurate approximation of the ground state energy per site, \cite{averagecorr} as listed in Table \ref{table:gsenergy}. In \fref{fig:correlators} we plot the 2-point correlator $C_2(r)$ of the energy density $h^{^{[i,i+1]}}$ [Eqs.~(\ref{eq:ising})-(\ref{eq:fib})] for the ground state $\ket{\Psi_{gs}}$,
\begin{equation}
C_2(r) = \langle h^{^{[i,i+1]}}h^{^{[i+r,i+r+1]}}\rangle - \langle h^{^{[i,i+1]}}\rangle \langle h^{^{[i+r,i+r+1]}}\rangle.\label{eq:2point}
\end{equation}
The expected polynomial decay $C_2(r) \propto 1/r^{x}$ is reproduced with exponents $x^{\tiny \mbox{Ising}} \approx 2.000$ and $x^{\tiny \mbox{Fib}} \approx 1.762$.
These exponents are compared with results from conformal field theory. In a CFT, 2-point correlators of a primary field $\phi_{h,\overline{h}}$ with conformal dimensions $(h,\overline{h})$ decay as
\begin{equation}
\langle \phi_{h,\overline{h}}(z,\overline{z}),~\phi_{h,\overline{h}}(0,0) \rangle = z^{^{-2h}}~\overline{z}^{^{-2\bar{h}}},
\end{equation}
where $z$ is the complex space-time coordinate and $\overline{z}$ is the conjugate of z (treated as an independent coordinate). The exponents $x^{\tiny \mbox{Ising}} \approx 2.000$ and $x^{\tiny \mbox{Fib}} \approx 1.762$ are consistent with the correlator \eeref{eq:2point} receiving dominant contribution from the energy density field $\epsilon$ ($h=\overline{h}=\frac{1}{2}$) of the Ising CFT, which predicts $x=2$, and from the spin field $\sigma$ ($h=\overline{h}=\frac{7}{16})$ of the Tricritical Ising CFT, which predicts $x=1.75$.

In \fref{fig:entropy} we plot the entanglement entropy
\begin{equation} 
S(r)=-\mbox{Tr}\left[\rho_r\log{}(\rho_r)\right],\label{eq:rho}
\end{equation}
of a block of $r$ anyons in the ground state $\ket{\Psi_{gs}}$, described by the reduced density matrix $\rho_r$. The expected logarithmic scaling $S(r) = (c/3)\log{}(r)$ for critical ground states is reproduced, and the central charges are approximated as $c^{\tiny \mbox{Ising}} \approx 0.5000[1]$ and $c^{\tiny \mbox{Fib}} \approx 0.70[1]$, in excellent agreement with the theoretical results of $1/2$ and $7/10$ respectively. 

\subsection{Anyonic Hubbard model\label{ssec:hubbard}} 

\begin{figure}
\begin{center}
\includegraphics[width=\columnwidth]{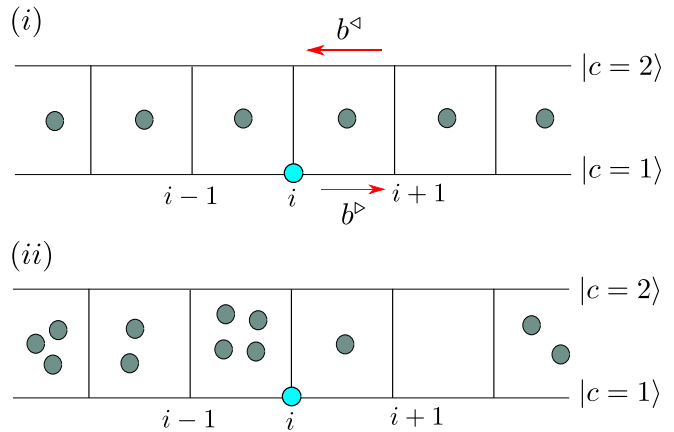}
\end{center}
\caption{The anyonic Hubbard model $(i)$ without topological disorder, and $(ii)$ with topological disorder corresponding to a uniform and a non-uniform filling of the islands of the ladder with pinned anyons respectively. One mobile anyon hops (here shown on site $i$) between vertices of the ladder around islands constaining pinned anyon(s). The Hamiltonian is the sum of a hopping term that corresponds to horizontal translation along the two legs of the ladder [which is associated with clockwise or counterclockwise braiding as described by operators $b^{\triangleright}$ and $b^{\triangleleft}$ respectively in \eref{eq:braidop}], and a tunnelling between legs with no braiding.}
\label{fig:ladder}
\end{figure}

We have also studied the dynamics of an anyonic Hubbard-like model. This model describes the hopping of mobile anyons on sites of a ladder with two horizontal legs, around islands which are occupied by pinned anyons (see \fref{fig:ladder}). The ladder is the minimal geometry which can accommodate interactions between anyons mediated purely via braiding, allowing mobile anyons on the ladder to braid around the pinned anyons. We consider a possibly non-translationally invariant (disordered) filling but restrict to the case of one mobile anyon hopping on the ladder. \cite{restrictHop} 

For a ladder made of $L$ sites, indexed by integer position $i$, the system can be described by a Hilbert space
\begin{equation}
\mathbb{V}^{\text{fusion}} \otimes (\mathbb{C}^3)^{\otimes L},
\end{equation}
where $\mathbb{V}^{\text{fusion}}$ is the fusion space of the pinned anyons plus the mobile anyon, and the $L$ qutrits $\ket{c_i}$ ($c \in \{0,1,2\}$) encode the position of the mobile anyon on the ladder: $\ket{0_i}$ corresponds to the absence of the mobile anyon on site $i$, $\ket{1_i}$ corresponds to the presence of the mobile anyon on the lower leg at site $i$ and $\ket{2_i}$ corresponds to the presence of the mobile anyon on the upper leg at site $i$.

The Hamiltonian is the sum of terms mediating hopping along the length of the ladder and terms mediating tunnelling between the upper and lower legs of the ladder,
\begin{equation}
H=H_{\text{hop}}+H_{\text{tun}}, \label{eq:hubbard}
\end{equation}
where
\begin{eqnarray}
H_{\text{hop}} &=& J\sum_{i=1}^{L-1}(T_i^+ b^{\triangleright}_i P_1+T_{i+1}^- b^{\triangleleft}_i P_2) + \text{h.c.},~J\in \mathbb{R},~~~\label{eq:Htrans} \\
H_{\text{tun}} &=& \mathbb{I}^{\text{fusion}} \otimes \sum_i(\kappa\ket{2_i}\bra{1_i}+\kappa^*\ket{1_i}\bra{2_i}),~\kappa\in \mathbb{C}\label{eq:Htun}.
\end{eqnarray}
($\mathbb{I}^{\text{fusion}}$ is the Identity on the fusion space of the anyons.) Here $T_i^\pm$ are translation operators between sites $i$ and $i\pm1$,
\begin{equation}
T_i^\pm = \mathbb{I}^{\text{fusion}} \otimes (\ket{1_{i\pm1}}\;\bra{1_i}+\ket{2_{i\pm1}}\;\bra{2_i}).
\end{equation}
On an open chain we assume reflecting boundary conditions: $T_1^- = T_L^+ = \mathbb{I}$. The operators $b^{\triangleright}_i$ and $b^{\triangleleft}_i$ braid the mobile anyon across the island immediately to the right of site $i$ and may be written as
\begin{align}
b^{\triangleright}_i &= b_{i,1} b_{i,2} \dotsm b_{i,m_i}, \\
b^{\triangleleft}_i  &= b_{i,m_i} b_{i,m_i-1} \dotsm b_{i,1},\label{eq:braidop}
\end{align}
where $m_i\in \mathbb{N}$ is the number of anyons in the island and the operators $\{b_{i,k}\}$ are a unitary representation of the $r$-strand braid group, $r=1+\sum_{i=1}^{n} m_i$ acting on the fusion space \cite{orderislands} of the anyons on the island immediately to the right of site $i$. Note, $b^{\triangleright}_i = b^{\triangleleft}_i = \mathbb{I}$ if $m_i=0$.
Finally, the projectors 
\begin{equation}
P_c= \mathbb{I}^{\text{fusion}} \otimes \sum_i\ket{c_i}\bra{c_i}
\end{equation}
act to select out states where the hopping anyon is on the lower or upper leg for $c=1$ and $c=2$ respectively.

In order to simulate this model using the anyonic TEBD algorithm we mapped the model on the ladder with $L$ sites to a one dimensional lattice $\mathcal{L}$ also with $L$ sites. Because the anyons on the islands are pinned, and the mobile anyon $a_{\text{mob}}$ braids around them \emph{en masse}, we may replace each island $i$ with a single anyon having the same total charge $a_{\text{pin}_i}$ as all the pinned anyons located in the island. Note that in the presence of disorder charge $a_{\text{pin}_i}$ can assume multiple values on some islands. We describe site $j$ of $\mathcal{L}$ by a basis labelled as
\begin{equation}
\ket{a_j,n_j,t_j},\label{eq:mydesc}
\end{equation}
where $a_j$ is an anyon charge, $n_j=0,1$ is a U(1) charge corresponding to the number of mobile anyons at site $j$ [see Appendix \ref{sec:general}] and $t_j=1,2$ labels the lower ($t_j=1$) or upper ($t_j=2$) leg of the ladder. The correspondence between the description of the model in terms of sites (\ref{eq:mydesc}) and the ladder system is as follows:
\begin{enumerate}
	\item $(a_j=a_{\text{mob}}, n_j=1,t_j=1) \Rightarrow$ Mobile anyon on lower leg at site $j$ of the ladder.
	\item $(a_j=a_{\text{mob}}, n_j=1,t_j=2) \Rightarrow$ Mobile anyon on upper leg at site $j$ of the ladder.
	\item $(a_j = a_{\text{pin}_j}, n_j=0, t_j=1) \Rightarrow$ Mobile anyon anywhere to the right of island $j$ of the ladder.
	\item $(a_j = a_{\text{pin}_{j-1}}, n_j=0, t_j=1) \Rightarrow$ Mobile anyon anywhere to the left of island $j$ of the ladder.
\end{enumerate}

We treat the pair $(a_j, n_j)$ as a \textit{composite} anyonic charge [see Appendix \ref{sec:general}] with degeneracy $t_{j}$, and describe lattice $\mathcal{L}$ by a fusion tree with $L$ open edges that are  labelled by composite charges. The fusion rules for the composite charges are given by
\begin{equation}
(a,n) \times (a',n') \rightarrow (a\times a',n+n').\label{eq:hubbardfuse}
\end{equation}
We truncate the U(1) charge $n+n'$ to a maximum value of 1, which imposes the constraint that the total number of mobile anyons on the ladder is equal to 1.

To illustrate this description, consider the anyonic Hubbard model that describes a single mobile Ising anyon with charge $a_{\text{mob}}=\sigma$ hopping on the ladder and with a single Ising anyon pinned in each island [\fref{fig:ladder}(i)], $a_{\text{pin}_j}=\sigma$. The basis on site $j$ is given by
\begin{equation}
\begin{split}
&\ket{a_j=\sigma,n_j=0,t_j=1},\\
&\ket{a_j=\sigma,n_j=1,t_j=1},\\
&\ket{a_j=\sigma,n_j=1,t_j=2}.
\end{split}
\end{equation}
The fusion space of 2 adjacent sites $(a_j,n_j) \times (a_{j+1},n_{j+1}) \rightarrow (a_p,n_p)$ is described by the basis
\begin{equation}
\begin{split}
&\ket{a_p=1,~n_p=0,~t_p=1},\\
&\ket{a_p=1,~n_p=1,~t_p\in\{1,2,3,4\}},\\
&\ket{a_p=\psi,~n_p=0,t_p=1},\\
&\ket{a_p=\psi,~n_p=1,t_p\in\{1,2,3,4\}}.
\end{split}
\end{equation}
The values of $t_p$ enumerate the possible configurations when one of the pair of anyons is the mobile anyon:
\begin{equation}
\begin{split}
t_p&=1 \Rightarrow \text{Mobile anyon on lower leg, left of island}.\\
t_p&=2 \Rightarrow \text{Mobile anyon on upper leg, left of island}.\\
t_p&=3 \Rightarrow \text{Mobile anyon on lower leg, right of island}.\\
t_p&=4 \Rightarrow \text{Mobile anyon on upper leg, right of island}.
\end{split} 
\end{equation}
In this description the Hamiltonian \eeref{eq:hubbard} can be expressed as the sum of two site terms $h^{^{[j,j+1]}}$ that are block diagonal, \cite{OBC}
\begin{equation}
\begin{split}
h^{^{[j,j+1]}} &= \bigoplus_{(a_p,n_p)} h^{^{[j,j+1]}}_{(a_p,n_p)}\\
 &= h^{^{[j,j+1]}}_{(1,0)} \oplus h^{^{[j,j+1]}}_{(1,1)} \oplus h^{^{[j,j+1]}}_{(\psi,0)} \oplus h^{^{[j,j+1]}}_{(\psi,1)},
\end{split}
\end{equation}
where
\begin{align}
\begin{split}
h^{^{[j,j+1]}}_{(1,0)} &\equiv (0), ~~~ h^{^{[j,j+1]}}_{(1,1)} \equiv \begin{pmatrix} 0 & \frac{\kappa^*}{2} & x & 0 \\ \frac{\kappa}{2} & 0 & 0 & x \\ x^* & 0 & 0 & \frac{\kappa^*}{2} \\ 0 & x^* & \frac{\kappa}{2} & 0 \end{pmatrix} , \\
h^{^{[j,j+1]}}_{(\psi,0)} &\equiv (0), ~~~ h^{^{[j,j+1]}}_{(\psi,1)} \equiv \begin{pmatrix} 0 & \frac{\kappa^*}{2} & y & 0 \\ \frac{\kappa}{2} & 0 & 0 & y \\ y^* & 0 & 0 & \frac{\kappa^*}{2} \\ 0 & y^* & \frac{\kappa}{2} & 0 \end{pmatrix}.
\end{split}\label{eq:hamnew}
\end{align} 
Here $x = R^{\sigma\sigma}_1 J$ and $y = R^{\sigma\sigma}_\psi J$, 
where $J$ and $\kappa$ are the couplings which appear in \eref{eq:hubbard} and $R^{ab}_c$ are the $R$-coefficients of the Ising anyon model [see \eref{eq:R} of Appendix~\ref{sec:appA}]. The anyonic Hubbard Hamitonian for a disorded filling of the islands can be described in a similar way. The physical states of the model are then states on the lattice $\mathcal{L}$ that have total anyon charge $a_\mathrm{tot}$ and total occupation $n_\mathrm{tot}=1$. These states can be represented as an anyonic MPS by replacing the anyon charges $a_j$ and $u_j$ that appear on the physical and bond indices in \fref{fig:anyonicmps} with composite charges $(a_j,n_j)$ and $(u_j,n_j)$ respectively.
\begin{figure}
\begin{center}
\includegraphics[width=7cm]{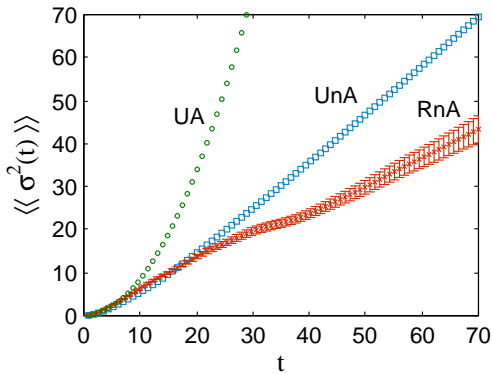}
\end{center}
\caption{Transport behaviour of anyons described by the anyonic Hubbard model, studied using the anyonic TEBD algorithm [with the constraint $\chi \leq 100$ in \eref{eq:chi00}]. The couplings in Eqs.~(\ref{eq:Htrans})--(\ref{eq:Htun}) are set as $J=\kappa=1$. The situation under study comprises a single anyon ($n_{\mathrm{tot}}=1$), initialised in the middle of a ladder with 100 sites, and permitted to hop around topologically charged islands filled with anyons of the same type (and with the total charge of all anyons set to $a_\mathrm{tot}=1$). Transport of this anyon along the chain is quantified by the variance of the spatial distribution $\langle \sigma^2(t)\rangle$. In the case of disordered topological backgrounds the variance is averaged over randomly distributed island charge occupations and is denoted $\langle\langle\sigma^2(t)\rangle\rangle$. Shown are the variances for (i) Abelian anyons with $\frac{\pi}{8}$ exchange statistics around a uniform background of islands  (Uniform Abelian=UA) showing ballistic transport, (ii) Ising anyons around a uniform background of Ising anyons (Uniform non-Abelian=UnA) exhibiting dispersive transport, and Ising model anyons with islands having random occupation levels $m_s\in\{0,\ldots,4\}$ of Ising anyons (Random non-Abelian=RnA; 50 configurations) also exhibiting dispersive transport but with a smaller diffusion constant (slope) than in the uniform non-Abelian case. For the random case we have also plotted the error bars for one sigma variance. The space and time axes are scaled so that the UnA case corresponds to a classical diffusion\cite{LZBPW,Lehman2013} with diffusion constant (slope) equal to one. Note that for the uniform cases we have $\langle\langle\sigma^2(t)\rangle\rangle = \langle \sigma^2(t)\rangle$. 
\label{fig:2}}
\end{figure}

We studied the real time dynamics of the anyonic Hubbard model using the anyonic TEBD algorithm. Our results are plotted in \fref{fig:2} and were presented in an earlier work. \cite{ZLSPB} In the case of uniform topological backgrounds, i.e.~translationally invariant filling of the islands, Abelian anyons have ballistic transport as indicated by the variance of the mobile anyons' spatial distribution, $\langle \sigma^2(t)\rangle\sim t^2$. In contrast non-Abelian anyons display a dispersive transport, $\langle \sigma^2(t)\rangle\sim t$, due to the fact that the different trajectories of the particle become correlated with different fusion environments while braiding, meaning spatial coherences are quickly lost. \cite{LZBPW,Lehman2013} In the presence of topological disorder, the behaviour changes substantially.  For Abelian anyons, the disorder acts to localize the particle while for non-Abelian anyons, the transport is still dispersive. This result is due to the fact that the fusion degrees of freedom become sufficiently entangled with the mobile anyon that the destructive interferences necessary to provide localisation are lost.  As described in detail in Ref. \onlinecite{ZLSPB} the competition between localization and decoherence is a subtle one, and long-time simulations using the anyonic MPS were essential to establish this result.

\subsection{Outlook} 
In this paper we have introduced the matrix product decomposition of states of 1D lattice systems of anyons, and have described how to extend the TEBD algorithm to the anyonic MPS. We have demonstrated the efficacy of the anyonic TEBD algorithm by computing the expected scaling of the ground state entanglement and 2-point correlators for two critical antiferromagnetically coupled infinite chains of non-Abelian anyons, our results being in agreement with those previously obtained
by authors using other techniques. Our method has the advantages that it is conceptually simple, demands only modest computational power to achieve accurate results, and can be applied to the study of generic anyon lattice models.

The basic data $(1-5)$ listed in Sec.~\ref{sec:intro} which characterize an anyon model may also be used to describe fermionic constraints or constraints due to the presence of an onsite global symmetry\cite{onsite} where charges correspond to the irreducible representations (irreps) of a symmetry group $\mathcal{G}$, as illustrated in Appendix~\ref{sec:appA}. By furnishing the data $(1-5)$ from a symmetry group $\mathcal{G}$ in this way, the anyonic MPS can also be used to efficiently represent states of a lattice system that are invariant, or more generally covariant, under the action of an onsite Abelian or non-Abelian global symmetry $\mathcal{G}$ on the lattice, as this is equivalent to a specialization from the more general anyonic case. Our implementation of anyonic constraints in the MPS is closely related to (and generalizes) the implementation of global onsite symmetry constraints in tensor network algorithms (see e.g.~Refs.~\onlinecite{Singh101,Singh102,Singh11,Singh12} and references therein). Similarly, the $\mathcal{G}-$symmetric MPS described in (for example) Ref.~\onlinecite{Singh101} provides an effective illustration of the present anyonic MPS formalism in what will be, for many, a more familiar context. 

We hope that the anyonic MPS formalism presented in this paper will prove to be a useful tool for studying generic lattice models of interacting anyons.

\acknowledgements

S.S. acknowledges financial support from the MQNS grant by Macquarie University Grant No. 9201200274. S.S. also thanks Robert Pfeifer and Perimeter Institute for hospitality. R.N.C.P. thanks the Ontario Ministry of Research and Innovation Early Researcher Awards for financial support.
G.K.B. thanks the KITP where part of this work was completed with support from the National Science Foundation under Grant No. NSF PHY11-25915.  This research was supported in part by the ARC Centre of Excellence in Engineered Quantum Systems (EQuS), Project No. CE110001013.  This research was supported in part by Perimeter Institute for Theoretical Physics. Research at Perimeter Institute is supported by the Government of Canada through Industry Canada and by the Province of Ontario through the Ministry of Research and Innovation.

\appendix
\section{Schmidt decomposition for spin systems\label{sec:appAA}}
In this Appendix we recapitulate the derivation of the Schmidt decomposition for spin systems as an aid for understanding the analogous derivation for the Schmidt decomposition of anyonic systems that is presented in Sec.\ref{ssec:bidecom}.

Consider a pure state $\ket{\Psi^{\mbox{\tiny [spin]}}}$ of a spin system that belongs to a tensor product space $\mathbb{V}^{(\mathcal{A})} \otimes \mathbb{V}^{(\mathcal{B})}$. State $\ket{\Psi^{\mbox{\tiny [spin]}}}$ can be expanded as
\begin{equation}
\ket{\Psi^{\mbox{\tiny [spin]}}} = \sum_{ab} \Psi^{\mbox{\tiny [spin]}}_{ab} \ket{a} \otimes \ket{b},\label{eq:spinbistate}
\end{equation}
where $\{\ket{a}\}$ and $\{\ket{b}\}$ denote an orthonormal basis in $\mathbb{V}^{(\mathcal{A})}$ and $\mathbb{V}^{(\mathcal{B})}$ respectively. We can regard $\Psi^{\mbox{\tiny [spin]}}$ as a matrix with components $\Psi^{\mbox{\tiny [spin]}}_{ab}$ where indices $a$ and $b$ enumerate the rows and columns. Consider the singular value decomposition of matrix $\Psi^{\mbox{\tiny [spin]}}$,
\begin{equation}
\Psi^{\mbox{\tiny [spin]}}_{ab} = \sum_{u}\Gamma^{^{[\mathcal{A}]}}_{au} \lambda_{u} \Gamma^{^{[\mathcal{B}]}}_{ub},\label{eq:spinsvd1}
\end{equation}
where $\Gamma^{^{[\mathcal{A}]}}$ and $\Gamma^{^{[\mathcal{B}]}}$ are unitary matrices,
\begin{equation}
\Gamma^{^{[\mathcal{A}]}} (\Gamma^{^{[\mathcal{A}]}})^\dagger = \Gamma^{^{[\mathcal{B}]}} (\Gamma^{^{[\mathcal{B}]}})^\dagger = \mathbb{I},
\end{equation}
and $\lambda$ is a diagonal matrix with non-negative diagonal entries $\lambda_{u} \geq 0$. Using (\ref{eq:spinsvd1}) in (\ref{eq:spinbistate}) we obtain
\begin{equation}
\ket{\Psi^{\mbox{\tiny [spin]}}} = \sum_{abu}\Gamma^{^{[\mathcal{A}]}}_{au} \lambda_{u} \Gamma^{^{[\mathcal{B}]}}_{ub} ~\ket{a} \otimes \ket{b}.
\end{equation}
Summing over $a$ and $b$ we obtain
\begin{equation}
\ket{\Psi^{\mbox{\tiny [spin]}}} = \sum_{u} \lambda_{u} ~\ket{\Phi^{^{\mathcal{A}}}_u} \otimes \ket{\Phi^{^{\mathcal{B}}}_u},\label{eq:spinsd11}
\end{equation}
where we have defined vectors
\begin{align}
\ket{\Phi^{^{\mathcal{A}}}_u} &= \sum_{a} \Gamma^{^{[\mathcal{A}]}}_{au} \ket{a},\\
\ket{\Phi^{^{\mathcal{B}}}_u} &= \sum_{a} \Gamma^{^{[\mathcal{B}]}}_{ub} \ket{b}.
\end{align}
By construction, vectors $\{\ket{\Phi^{^{\mathcal{A}}}_u}\}$ and $\ket{\Phi^{^{\mathcal{B}}}_u}$ are orthonormal,
\begin{equation}
\braket{\Phi^{^{\mathcal{A}}}_u}{\Phi^{^{\mathcal{A}}}_{u'}} = \braket{\Phi^{^{\mathcal{B}}}_u}{\Phi^{^{\mathcal{B}}}_{u'}} = \delta_{uu'}.
\end{equation}
Equation (\ref{eq:spinsd11}) is the Schmidt decomposition of the bipartite state $\ket{\Psi^{\mbox{\tiny [spin]}}} \in \mathbb{V}^{(\mathcal{A})} \otimes \mathbb{V}^{(\mathcal{B})}$.

The norm of state $\Psi^{\mbox{\tiny [spin]}}$ is
\begin{equation}
\braket{\Psi^{\mbox{\tiny [spin]}}}{\Psi^{\mbox{\tiny [spin]}}} = \sum_u \lambda^{^2}_u.
\end{equation}
The Schmidt decomposition is a useful tool in quantum information theory to study bipartite entanglement. The reduced density matrices $\rho^{^{[\mathcal{A}]}}$ and $\rho^{^{[\mathcal{B}]}}$ for parts $\mathcal{A}$ and $\mathcal{B}$ of state $\ket{\Psi^{\mbox{\tiny [spin]}}}$ in (\ref{eq:spinsd11}) are obtained as
\begin{align}
\rho^{^{[\mathcal{A}]}} &= \sum_u \lambda^{^2}_u \ket{\Phi^{^{[\mathcal{A}]}}_u}\bra{\Phi^{^{[\mathcal{A}]}}_u},\\
\rho^{^{[\mathcal{B}]}} &= \sum_u \lambda^{^2}_u \ket{\Phi^{^{[\mathcal{B}]}}_u}\bra{\Phi^{^{[\mathcal{B}]}}_u}.
\end{align}
The von-neumman entanglement entropy $S(\mathcal{A}:\mathcal{B})$,
\begin{equation}
S(\mathcal{A}:\mathcal{B}) = -\mbox{Tr}(\rho^{^{[\mathcal{A}]}}\mbox{log}\rho^{^{[\mathcal{A}]}}) = -\mbox{Tr}(\rho^{^{[\mathcal{B}]}}\mbox{log}\rho^{^{[\mathcal{B}]}}),
\end{equation} 
of the bipartite state $\ket{\Psi^{\mbox{\tiny [spin]}}}$ is obtained as
\begin{equation}
S(\mathcal{A}:\mathcal{B}) = -\sum_u \lambda^{^2}_u \mbox{log}~\lambda^{^2}_u.
\end{equation}

\section{Anyon models\label{sec:appA}}
In this Appendix we introduce basic terminology and graphical notation pertaining to anyon models as used in this paper.
For those already familiar with graphical notations for states of anyonic systems, the formalism employed in this paper corresponds to that described in Ref.~\onlinecite{BondersonThesis}, 
save that it has been 
rotated 135$^\circ$ counterclockwise in order to emphasise the relationship between the tensor network structure of the MPS (\fref{fig:intro1}) and the corresponding anyonic fusion tree (\fref{fig:intro2}). The decision was made to rotate the fusion tree to match the tensor network, rather than rotating the tensor network to match the fusion tree as in \cite{}, because the target audience of this paper is primarily intended to be readers with prior experience in conducting simulations using MPS and DRMG, and thus it was considered desirable to reflect the familiar tensor network configuration as closely as possible using the anyonic model. For readers who are not familiar with the graphical notation for anyonic systems, we summarize the pertinent features below.

An anyon model consists of a finite set $\Omega$ of particle types, or charges, $a,b,c,\ldots$. The set $\Omega$ of allowed charges includes a distinguished trivial or vacuum charge $1 \in \Omega$.
\subsection{Two anyons}
The local properties of a single anyon are completely specified by its ``charge''. However, two (or more) anyons with charges $a$ and $b$ can be \textit{fused} together into a total charge $c$ which can, in general, take several values,
\begin{equation}
a\times b \rightarrow \sum_c N_{ab}^{c}~c.\label{eq:fusionrules}
\end{equation}
Here $N_{ab}^c$ is the number of times (or the \textit{multiplicity}) charge $c$ appears in the fusion outcome. We says charges $a,b$ and $c$ are \textit{compatible} with one another if $N_{ab}^c \neq 0$, and that tensor $N$ encodes the \textit{fusion rules} of the anyon model. For simplicity, we consider anyon models that are multiplicity free, namely, $N_{ab}^{c} = 0,1$ for all charges $a$ and $b$. However, multiplicities can be accommodated into our formalism in a rather straightforward way, as described in Appendix~\ref{sec:general}. The fusion rules for the vacuum charge satisfy $N_{a1}^b = N_{1a}^b = \delta_{ab}~\forall~a,b \in \Omega$. Two charges $a$ and $b$ are said to be \textit{dual} to one another, denoted as $a^* = b$ and $b^* = a$, if they fuse together to the vacuum. The operation $*:\Omega \rightarrow \Omega$ is an involution, $(a^*)^* = a$.

\begin{figure}
\includegraphics[width=6cm]{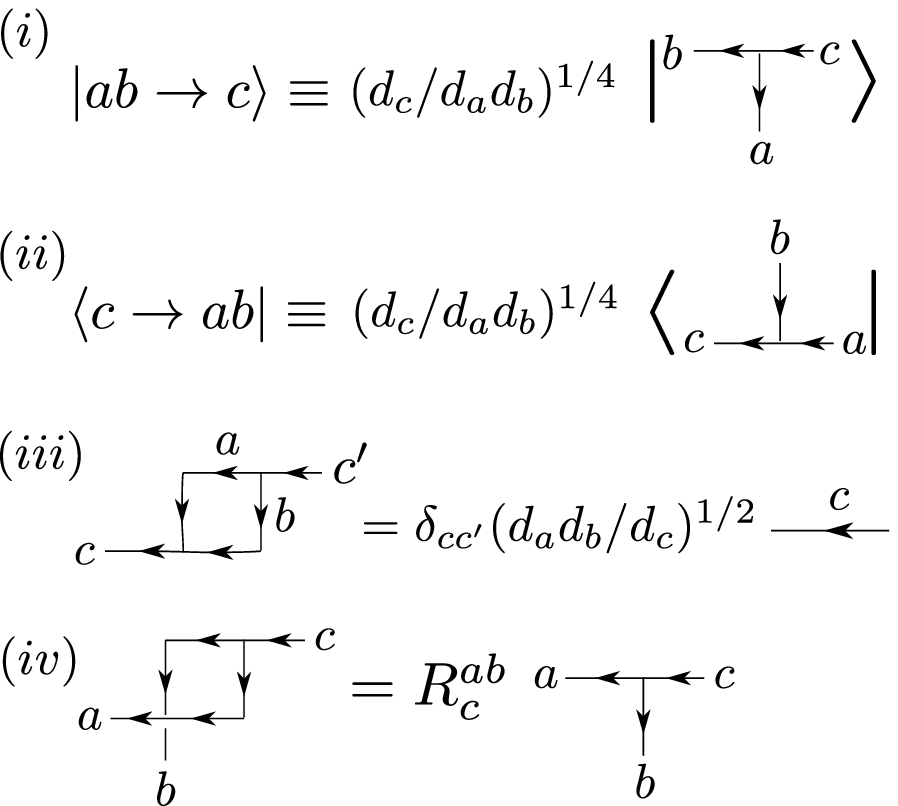}
\caption{The graphical representation of $(i)$ the ket $\ket{\fuse{a}{b}{c}}$, $(ii)$ the corresponding bra $\bra{c\rightarrow ab}$, $(iii)$ the inner product (braket), and $(iv)$ braiding two anyons. \label{fig:twoanyons}} 
\end{figure}

Consider two anyons $a$ and $b$ that are fixed on a line. We denote an orthonormal basis in the total Hilbert space as
\begin{equation}
\ket{\fuse{a}{b}{c}},
\end{equation}
where charge $c$ is obtained by fusing $a$ and $b$. The graphical representations of the ket $\ket{\fuse{a}{b}{c}}$ and the corresponding bra $\bra{\fuse{a}{b}{c}}$ as employed in this paper are shown in \fref{fig:twoanyons}(i)-(ii). Their inner product $\braket{c'\rightarrow a b}{\fuse{a}{b}{c}}$ is graphically represented by gluing the diagrams of the ket and bra as shown in \fref{fig:twoanyons}(iii). 

The ordering of anyons $a$ and $b$ on the line may be interchanged by braiding them around one another to obtain another basis $\ket{\fuse{b}{a}{c}}$. We follow the convention that when braiding $a$ counterclockwise around $b$, $\ket{\fuse{b}{a}{c}}$ is related to $\ket{\fuse{a}{b}{c}}$ by a 3-index tensor $R$ [see \fref{fig:twoanyons}(iv)],
\begin{equation}
\ket{\fuse{a}{b}{c}} = R^{ab}_c \ket{\fuse{b}{a}{c}},\label{eq:R}
\end{equation}
while braiding $a$ clockwise around $b$ relates the two bases as
\begin{equation}
\ket{\fuse{a}{b}{c}} = (R^{ab}_c)^* \ket{\fuse{b}{a}{c}},
\end{equation}
where * denotes complex conjugation.

\subsection{Three anyons}
\begin{figure}
\includegraphics[width=\columnwidth]{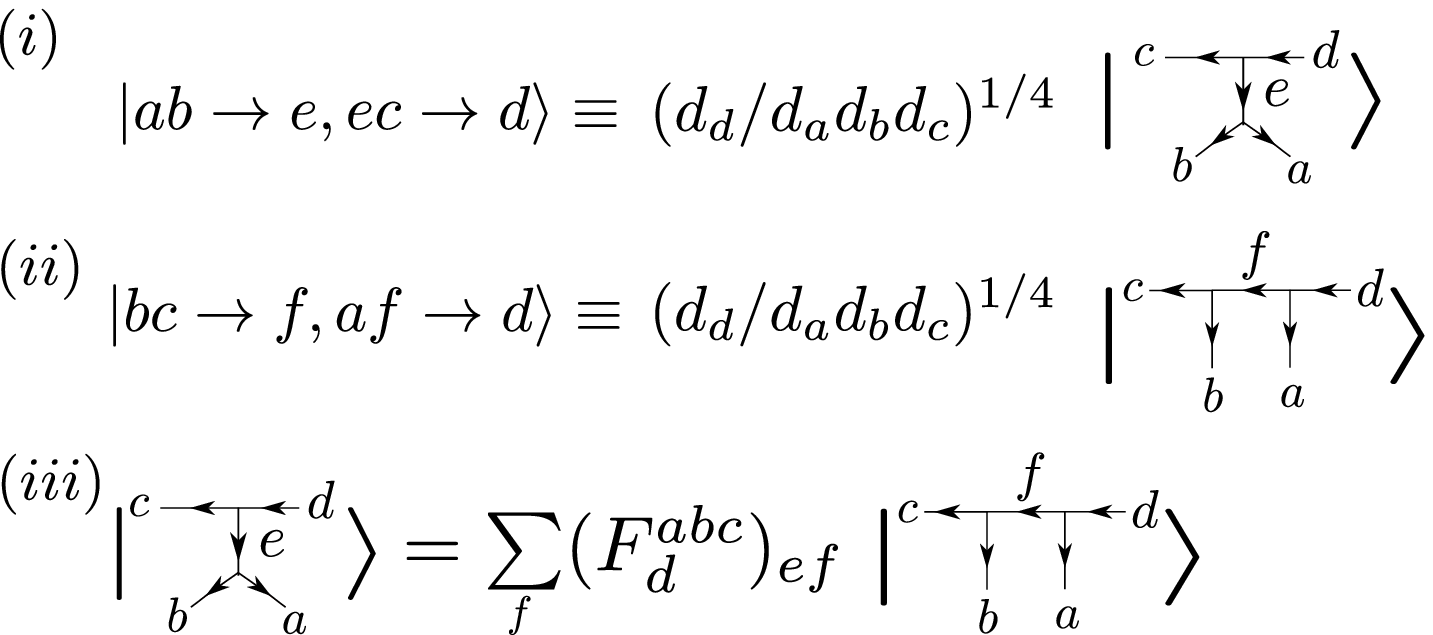}
\caption{$(i)-(ii)$ Two possible fusion bases for three anyons $a,b$ and $c$. $(iii)$ $F$-move, the unitary transformation which implements the change of basis from (i) to (ii). \label{fig:threeanyons}} 
\end{figure}

For three (or more) anyons fixed on a line, different choices of basis are possible corresponding to different ways of performing pairwise fusings of the anyons into a single total charge. Three anyons $a,b$ and $c$ can be fused to a total charge $d$ by first fusing $a$ and $b$ into an intermediate charge $e$ and then fusing $e$ and $c$ to total charge $d$. Denote the corresponding basis by $\ket{\fuse{a}{b}{e},\fuse{e}{c}{d}}$. Alternatively,  we could first fuse $b$ and $c$ into $f$ and then fuse $a$ and $f$ into $d$. Denote the basis corresponding to this fusion sequence by $\ket{\fuse{b}{c}{f},\fuse{a}{f}{d}}$. The two bases are related by a unitary transformation given by a 6-index tensor $F$ (see \fref{fig:threeanyons}),
\begin{equation}
\ket{\fuse{a}{b}{e},\fuse{e}{c}{d}} = \sum_f (F^{abc}_d)_{ef} \ket{\fuse{b}{c}{f},\fuse{a}{f}{d}}.\label{eq:F}
\end{equation}
The transformation given in \eref{eq:F} is also known as an \textit{F-move}.

\subsection{Arbitrary number of anyons} 
Let us now consider a one dimensional lattice $\mathcal{L}$ made of $L$ sites that are fixed on a line and populated by anyons $\{a_i\}_{i=1}^L$ that belong to a given anyon model. Denote by $\mathbb{V}^{(\mathcal{L})}$ the Hilbert space that describes lattice $\mathcal{L}$. Generalizing the description for three anyons, we introduce an orthonormal basis in $\veclat{L}$ by means of a \textit{fusion tree}. A fusion tree corresponds to a particular sequence of pairwise fusions of the $L$ anyons $\{a_i\}_{i=1}^L$ into a total definite charge $a_\mathrm{tot}$ by means of $L-1$ intermediate charges $\{u_i\}_{i=1}^{L-1}$. For example, a possible choice of fusion tree for a lattice made of $L=5$ sites is shown in \fref{fig:intro2}. Different choices of fusion trees correspond to different choice of bases, and are related one another by $F$-moves. When each site of the lattice carries the same charge $a$, the dimension of the Hilbert space $\mathbb{V}^{(\mathcal{L})}$ is found to scale as 
\begin{equation}
\lim_{L\rightarrow \infty}\left|\mathbb{V}^{(\mathcal{L})}\right|\longrightarrow (d_a)^L 
\end{equation}
where $d_a$ is the \textit{quantum dimension} of charge $a$, being analogous to the dimension of an irreducible representation of a group. In general this expression only holds approximately for finite $L$, because $d_a$ may be non-integer for anyon models.

\subsection{Examples of anyon models\label{ssec:basicdata}}
An anyon model is completely specified by the following set of data:
\begin{enumerate}
\item The set of allowed charges, $\Omega$.
\item The quantum dimension $d_a$ of each charge $a$.
\item The fusion rules, encoded in the multiplicity tensor $N_{ab}^c$ of \eref{eq:fusionrules}.
\item The braiding coefficients $R^{ab}_c$ of \eref{eq:R}.
\item The $F$-move coefficients $(F^{abc}_d)_{ef}$ of \eref{eq:F}.
\end{enumerate}
For a consistent anyon model the $F$ and $R$ coefficients are required to satisfy the \textit{pentagon} and \textit{hexagon} relations that express associativity of fusion and compatibility of fusion with braiding respectively. \cite{BondersonThesis} A broad class of anyon models is described by the quantum symmetry groups SU(2)$_k,~~k \in \mathbb{Z}^+$ where the set of allowed charges corresponds to the irreducible representations $\Omega \equiv \{0,1/2,1,\ldots,k/2\}$ of SU(2)$_k$ and, for instance, the $F$ coefficients correspond to the quantum 6-j symbols of the group. Next, we list some simple examples of anyon models. For a more extensive list of anyon models see also Ref.~\onlinecite{BondersonThesis}.

\subsubsection{Fibonacci anyon model\label{ssec:fib}}
The Fibonacci anyon model consists of two charges: $1$ (the vacuum) and $\tau$. The quantum dimensions of these charges are
\begin{equation}
d_1 = 1,~~~d_{\tau} = (1+\sqrt{5})/2.
\end{equation}
The only non-trivial fusion rule is $\tau\times\tau\rightarrow 1+\tau$. That is, all components $N_{ab}^c$ are zero except
\begin{align}
N_{\tau \tau}^{1} = N_{\tau \tau}^{\tau} = N_{\tau 1}^{\tau} = N_{1 \tau }^{1} = N_{1 1}^{1} = 1.\nonumber
\end{align}
The coefficients $R^{ab}_c$ are non-zero only if $N_{ab}^c = 1$. The non-zero $R$-coefficients are
\begin{align}
R^{\tau \tau}_{1} = e^{-i4\pi/5},~R^{\tau \tau}_{\tau} = e^{i3\pi/5},~R^{1 \tau}_{1} = R^{1 \tau}_{1} = R^{1 1}_{1} = 1.\nonumber
\end{align}
The non-trivial $F$-move coefficients are
\begin{align}
(F^{\tau\tau\tau}_{\tau})_{ef} = \begin{pmatrix} \phi^{-1} & \phi^{-1/2} \\ \phi^{-1/2} & -\phi^{-1} \end{pmatrix},\nonumber
\end{align}
where $\phi = (1+\sqrt{5})/2$ and $e,f \in \{1,\tau\}$. The remaining $F$-move coefficients are given by
\begin{equation}
(F^{abc}_{d})_{ef} = N_{ab}^e N_{bc}^f N_{ec}^d 
N_{af}^d.\label{eq:trivialF}
\end{equation}

\subsubsection{Ising anyon model\label{ssec:ising}}
The Ising anyon model consists of three charges, $1,\sigma$ and $\psi$, with quantum dimensions
\begin{equation}
d_1 = d_{\psi} = 1,~~~d_{\sigma} = \sqrt{2}.
\end{equation}
The non-trivial entries in the multiplicity tensor are given by
\begin{align}
N_{\sigma \sigma}^{1} = N_{\sigma \sigma}^{\psi} = N_{\psi \sigma}^{\psi} = N_{\sigma \psi}^{\psi} = 1.\nonumber
\end{align}
The non-trivial $R$-coefficients are
\begin{align}
R^{\tau \tau}_{1} &= e^{-i\pi/8}, R^{\tau \tau}_{1} = e^{i3\pi/8}, R^{ \tau \psi}_{\tau} = R^{\psi \tau}_{\tau} = e^{-i\pi/2}, \nonumber \\ 
R^{\psi \psi}_{1} &= -1.\nonumber
\end{align}
The non-trivial $F$-move coefficients are
\begin{align}
(F^{\sigma\sigma\sigma}_{\sigma})_{ef} = \begin{pmatrix} \frac{1}{\sqrt{2}} & \frac{1}{\sqrt{2}} \\ \frac{1}{\sqrt{2}} & -\frac{1}{\sqrt{2}} \end{pmatrix},~~~(F^{\sigma\psi\sigma}_{\psi})_{\sigma\sigma} = (F^{\psi\sigma\psi}_{\sigma})_{\sigma\sigma} = -1,\nonumber
\end{align}
where $e,f \in \{1,\psi\}$. The remaining $F$-move coefficients are once again given by \eref{eq:trivialF}.

\subsubsection{Fermions}
The relevant charge for fermions is the parity $p$ of fermion particle number. Charge $p$ takes two values, $p=0$ and $p=1$ corresponding to an even or odd number of fermions respectively. The fusion rules $N_{pp'}^{p''}$ are given by
\begin{equation}
N_{00}^0 = N_{01}^1 = N_{10}^1 = N_{11}^1 = 1,
\end{equation}
and $N_{pp'}^{p''}=0$ for all remaining values of $p,p'$ and $p''$.\\
The non-trivial $R$-coefficients are given by
\begin{equation}
R^{00}_0 = R^{01}_1 = R^{10}_1 = 1,~~R^{11}_0 = -1.
\end{equation}
The $F$-move coefficients are given by
\begin{equation}
(F^{abc}_{d})_{ef} = N_{ab}^e N_{bc}^f N_{ec}^d 
N_{af}^d.
\end{equation}

\section{Implementation of the anyonic TEBD algorithm\label{sec:atebd}}
In this Appendix we explain the step by step implementation of the main update [\fref{fig:mainstep}] of the anyonic TEBD algorithm.
\begin{figure}
\includegraphics[width=\columnwidth]{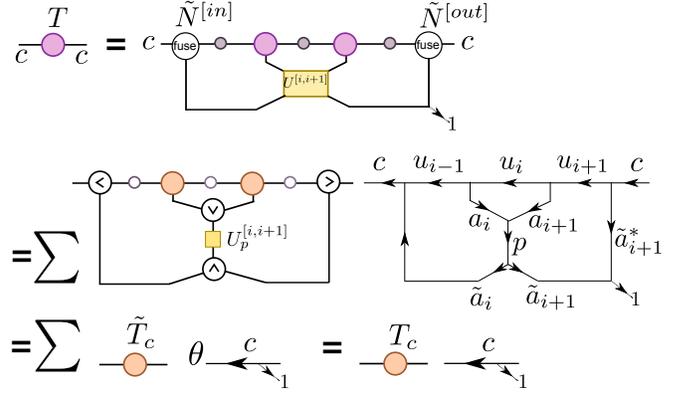}
\caption{Step 1 of the main update of the anyonic TEBD algorithm. Description in text. \label{fig:step21}} 
\end{figure}
\begin{figure}
\includegraphics[width=7cm]{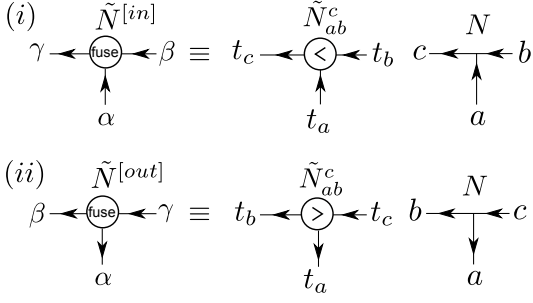}
\caption{The graphical representation of the \textit{generalized fusion tensor} $\tilde{N}$ that describes fusion $\alpha \times \beta \rightarrow \gamma$ of indices $\alpha$ and $\beta$ that carry both charge and degeneracy, $\alpha = (a,t_a)$, $\beta = (b,t_b)$ and $\gamma = (c,t_c)$. Here we show the generalized fusion tensor $\tilde{N}^{\tiny \mbox{[in]}}$  ($\tilde{N}^{\tiny \mbox{[out]}}$) that corresponds to the fusion of two incoming (outgoing) indices $\alpha$ and $\beta$ into an outgoing (incoming) index $\gamma$. The fusion tensors $\tilde{N}^{\tiny \mbox{[in]}}$ and $\tilde{N}^{\tiny \mbox{[out]}}$ satisfy $\tilde{N}^{\tiny \mbox{[in]}} \tilde{N}^{\tiny \mbox{[out]}} = \mathbb{I}$ when summed over $\alpha$ and $\beta$. 
For fixed charges $a, b$ and $c$, $\tilde{N}^{\tiny \mbox{[in,out]}}$ decomposes [\eref{eq:ndecompose}] into a degeneracy tensor $\tilde{N}_{ab}^{c}$ with components $(\tilde{N}_{ab}^c)_{t_at_b}^{t_c} = 0,1$ and a usual fusion tensor $N_{ab}^c$. In the graphical representation of degeneracy tensor $\tilde{N}_{ab}^{c}$ the caret points to the total charge, that is, when fusing incoming (outgoing) indices the caret points to the outgoing (incoming) index.\label{fig:ntensor}}
\end{figure}

First we introduce an important transformation -- the \textit{generalized fusion tensor} -- that is required in the algorithm. A generalized fusion tensor $\tilde{N}_{\alpha\beta}^{\gamma}$ describes fusion of indices
\begin{equation*}
\alpha \times \beta \rightarrow \gamma
\end{equation*}
that carry both charge and degeneracy, $\alpha = (a,t_a), \beta = (b,t_b)$ and $\gamma = (c,t_c)$, and thus generalizes the usual fusion tensor $N_{ab}^c$ that is defined for fusing charges,
\begin{equation*}
a \times b \rightarrow c
\end{equation*}
of an anyon model according to \eref{eq:fusionrules}. Components $\tilde{N}_{\alpha\beta}^{\gamma}$ are identically zero if charges $a$, $b$, and $c$ are incompatible with the fusion rules, i.e.~when $N_{ab}^c=0$. Denote by $n_a,n_b$ and $n_c$ the degeneracy of charges $a,b$ and $c$ respectively, namely,
\begin{align}
&t_a \in \{1,2,\ldots,n_a\},\\
&t_b \in \{1,2,\ldots,n_b\},\\
&t_c \in \{1,2,\ldots,n_c\}.
\end{align}
In general there exist multiple choices of $a$ and $b$ compatible with any given charge $c$. The total degeneracy $n_c$ of charge $c$ then corresponds to the total number of different ways by which charge $c$ may be obtained by fusing charges $a$ and $b$, with each degenerate copy of $a$ and $b$ counting as a separate channel. Thus
\begin{equation}
n_c=\sum_{a,b}N^c_{ab}n_an_b.
\end{equation}
For given values of $a$ and $b$, each value of $t_c$ can be associated to a pair $(t_a,t_b)$ so as to construct a one-to-one correspondence. We encode this association by setting the relevant component $\tilde{N}_{\alpha\beta}^{\gamma} = 1$. Thus, for fixed charges $a,b$ and $c$, tensor $\tilde{N}$ can be decomposed as
\begin{equation}
\tilde{N} = \!\!\bigoplus_{abc:N_{ab}^c}\!\! \tilde{N}_{ab}^{c},\label{eq:ndecompose}
\end{equation}
where $\tilde{N}_{ab}^{c}$ is a (degeneracy) tensor made of components $(\tilde{N}_{ab}^{c})_{t_at_b}^{t_c}=0,1$ that encodes the contributions of $(a,b)$ to the degeneracy of $c$. The graphical representation of the generalized fusion tensor and the decomposition (\ref{eq:ndecompose}) is shown in \fref{fig:ntensor}.

Next we explain how the update depicted in \fref{fig:mainstep} is performed in 4 steps.

\textit{Step 1} of the update is depicted in \fref{fig:step21}. It corresponds to the reduction of a section of the MPS to a charge-conserving matrix $T$,
\begin{equation}
T=\bigoplus_c T_c,\label{eq:matrixT}
\end{equation}
obtained by fusing indices and contracting tensors together as depicted at the top of \fref{fig:step21}.

Since indices carry both charge and degeneracy, we employ generalized fusion tensors $\tilde{N}^{\tiny \mbox{[in]}}$ and $\tilde{N}^{\tiny \mbox{[out]}}$ [\fref{fig:ntensor}] to fuse indices as shown in \fref{fig:step21}. A fusion vertex $\tilde{a}_{i+1} \times \tilde{a}^*_{i+1} \rightarrow 1$ is inserted to reverse the orientation of $\tilde{a}_{i+1}$ prior to fusion with $u_{i+1}$. The aforementioned fusion of indices and the index reversal is undone in Step 3. This can be understood as employing a resolution of Identity, $\mathbb{I} = YY^{-1}$ where, for example, transformation $Y$ comprised of the reversal of $\tilde{a}_{i+1}$ and the fusion of charges $u_{i+1}$ and $\tilde{a}^*_{i+1}$ is enacted in this step and the corresponding $Y^{-1}$ is enacted in Step 3. 

Each block $T_c$ of $T$ is determined independently by fixing charge $c$ on the open indices and summing over all compatible internal charges and their respective degeneracy indices. For fixed compatible internal charges, the tensors of the MPS decompose into degeneracy and fusion parts according to \eref{eq:gamma}, and as depicted in \fref{fig:anyonicmps}. By construction the generalized fusion tensors $\tilde{N}^{\tiny \mbox{[in]}}$ and $\tilde{N}^{\tiny \mbox{[out]}}$ also decompose into degeneracy and fusion parts as shown in \fref{fig:ntensor}. Thus, for fixed charges on all indices the entire contraction decomposes into two parts as shown in \fref{fig:step21}, namely, a part made of only degeneracy tensors and a fusion network made of only fusion tensors. The degeneracy tensors are contracted to obtain a matrix $\tilde{T}_c$. The fusion network encodes the fusion contraints on the entire contraction, and vanishes when some of the charges are incompatible with fusion rules. For compatible charges the fusion network can be replaced by a single fusion vertex $c \times 1 \rightarrow c$ multiplied by a factor $\theta$,
\begin{equation}
\theta = (F^{u_{i-1} a_i a_{i+1}}_{u_{i+1}})_{u_i p}~.~(F^{ \tilde{a}_{i+1} \tilde{a}_i u_{i-1}}_{u_{i+1}})_{p c}~.~(F^{c \tilde{a}_{i+1} \tilde{a}_{i+1}^*}_{c})_{u_{i+1} 1},
\end{equation}
by applying $F$-moves (\fref{fig:threeanyons}) and contracting loops [\fref{fig:twoanyons}(iii)]. 

Matrix $\tilde{T}_c$ is then multiplied by the factor $\theta$. The above procedure is repeated for all compatible internal charges, and the matrices $\tilde{T}_c$ thus obtained are summed together to obtain the total (degeneracy) block $T_c$. By iterating over all different charges $c$, all blocks $T_c$ of matrix $T$ are determined.

Figure \ref{fig:step22} depicts the remaining steps 2-4 of the local update. The figure shows each step both in terms of the full charge-conserving tensors (in the left column of each box) and as it is actually performed by decomposing tensors into degeneracy and fusion parts (in the right column of each box). Addressing each in turn:
\begin{figure}
\includegraphics[width=8.75cm]{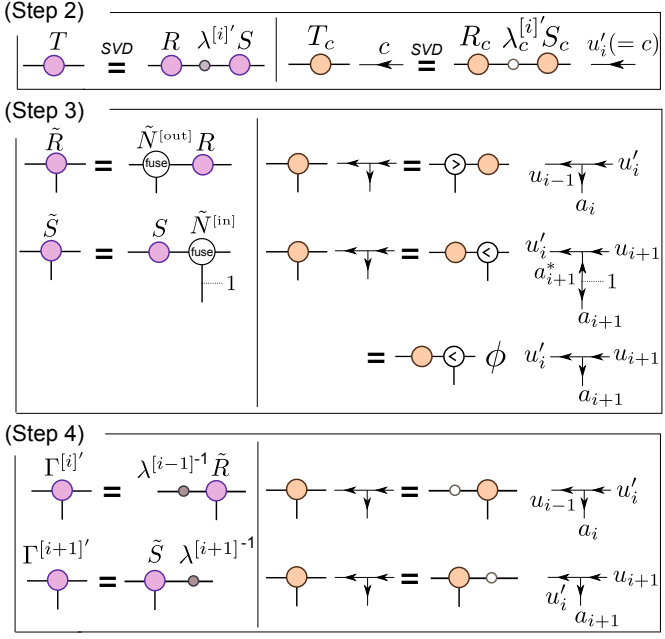}
\caption{Steps 2-4 of the main update of the anyonic TEBD algorithm. Description in text.\label{fig:step22}} 
\end{figure}

\textit{Step 2} is to singular value decompose matrix $T$ that is obtained at the end of Step 1. This can be achieved by singular value decomposing individual blocks $T_c$ of $T$ [\eref{eq:matrixT}],
\begin{equation}
T_c = R_c {\lambda^{^{[i]}}_c}' S_c.
\end{equation}
The matrix ${\lambda^{^{[i]}}}'$ and the index $\mu'_i = (c,t_c)$ thus obtained replace ${\lambda^{^{[i]}}}$ and index $\mu_i$ in the updated MPS respectively. In practical simulations, a truncation is made after the singular value decomposition by retaining only a fixed number $\chi$ of singular values ${\lambda^{^{[i]}}_c}'$. The truncation results in an updated index $\mu'_i = (c,t_{c})$ such that the norm,
\begin{equation}
\braket{\Psi}{\Psi} = \sum_{c}d_{c}\left[\sum_{t_{c}} ({\lambda^{^{[i]}}_{c}}')_{t_{c}}^{^2}\right],~~~ \sum_{c} |t_{c}| = \chi,\label{eq:chi}
\end{equation}
of the updated MPS $\ket{\Psi}$ is maximized, where $d_{c}$ is the quantum dimension of the charge $c$. If the norm of the state if to be held at 1, this may be achieved by rescaling the values ${\lambda^{^{[i]}}_{c}}'$ after truncation.

\textit{Step 3} is to reorganize matrices $R$ and $S$ into three-index tensors $\tilde{R}$ and $\tilde{S}$ respectively. This is achieved by contracting $R$ and $S$ with generalized fusion tensors $\tilde{N}^{\tiny \mbox{[out]}}$ and $\tilde{N}^{\tiny \mbox{[in]}}$ respectively as shown. These contractions also proceed as before by fixing charges on the indices, decomposing the tensors into degeneracy and fusion parts and then contracting the degeneracy tensors.

The reorganization of matrix $S$ also involves inserting a fusion vertex $\tilde{a}^*_{i+1} \times \tilde{a}_{i+1} \rightarrow 1$ to undo the reversal of $a_{i+1}$ enacted in Step 1 and to re-orient index $a_{i+1}$ as outgoing on the reorganized tensor $\tilde{S}$. Then by applying an $F$-move the two fusion vertices can be replaced by a single fusion vertex multiplied by a factor $\phi$,
\begin{equation}
\phi = (F^{a_{i+1} a^*_{i+1} u_{i+1}}_{u_{i+1}})_{1 u'_i}.
\end{equation}
This factor $\phi$ is multiplied into the reorganized tensor $\tilde{S}$.

\textit{Step 4} is to contract tensors $\tilde{R}$ and $\tilde{S}$ with the inverse matrices $\left({\lambda^{^{[i-1]}}}\right)^{-1}$ and $\left({\lambda^{^{[i+1]}}}\right)^{-1}$ as shown to obtain the updated tensors ${\Gamma^{^{[i]}}}'$ and ${\Gamma^{^{[i+1]}}}'$ and restore the canonical form of the anyonic MPS.

Finally, we make some remarks pertaining to a practical software implementation of the anyonic TEBD algorithm. First, for an anyon model with a finite number of charges, there are only a finite number of fusion networks that appear in Step 1 of the anyonic TEBD algorithm. These fusion networks depend only on the anyon charges and can therefore be enumerated and the factors $\theta$ [\fref{fig:step21}] may be computed \textit{prior} to executing the algorithm for the given Hamiltonian. This saves the CPU time that is otherwise expended to compute the same factors $\theta$ in each iteration of the TEBD algorithm. Second, each block $T_c$ is determined \textit{independently} by fixing different values of charge $c$ on the open indices and summing over all compatible internal charges. This fact can be exploited in the software implementation to parallelize the contractions in Step 1 that correspond to different values of $c$.

\section{Generalizations of the anyonic MPS formalism\label{sec:general}}
In this Appendix we discuss some generalizations of the anyonic MPS formalism that is presented in this paper. The formalism can be readily extended
\begin{enumerate}
\item to study lattice models with global onsite symmetry (Appendix~\ref{ssec:symmetrygroup}), and 
\item to study anyon lattice models with 
\begin{enumerate}
\item auxiliary degrees of freedom on the lattice sites (Appendix~\ref{ssec:supplementary}), and/or 
\item fusion multiplicities (Appendix~\ref{ssec:fusionmult}).
\end{enumerate}
\end{enumerate}

\subsection{Onsite global symmetries\label{ssec:symmetrygroup}}
The basic data $(1-5)$, listed in the previous section, which characterize an anyon model may also be used to describe properties of a regular symmetry group $\mathcal{G}$ where the charges correspond to the irreps of $\mathcal{G}$.

\textit{Example 1:} Consider an Abelian Lie group $\mathcal{G} =$ U(1). Charges (irreps) of U(1) are labelled by integers $n$ and have dimension 1. The fusion rules are
\begin{equation}
n_1 \times n_2 \rightarrow n = n_1+n_2.
\end{equation}
The $R$ and $F$ coefficients are trivial for U(1) (and for any Abelian group). That is, all $R$ and $F$ coefficients are equal to 1 for compatible charges and equal to 0 otherwise. 

\textit{Example 2:} Consider the non-Abelian group $\mathcal{G} =$ SU(2). Charges of SU(2) are labelled by non-negative semi-integers $n=0,1/2,1,3/2,\ldots$. The dimension of charge $n$ is equal to $2n+1$. The fusion rules are
\begin{equation}
n_1 \times n_2 \rightarrow \{n:|n_1-n_2|,|n_1-n_2|+1,\ldots,n_1+n_2\}.
\end{equation}
The $R$-coefficients are given by
\begin{equation}
R^{n_1n_2}_n = (-1)^{n_1+n_2-n}.
\end{equation}
The $F$-move coefficients correspond to the 6-j symbols of the SU(2).

By furnishing the data $(1-5)$ from a symmetry group $\mathcal{G}$, the anyonic MPS can be used to represent states of a lattice system that are invariant or, more generally, covariant under the action of an onsite global symmetry $\mathcal{G}$ on the lattice. (An onsite global symmetry means that the symmetry group acts identically on each site of the lattice.) For example, in the context of lattice spin systems, a global onsite symmetry $\mathcal{G}=\mathrm{SU(2)}$  may correspond to invariance of total spin under an identical rotation of all spins. An anyonic MPS constructed from the fusion data furnished from $\mathcal{G}=$ SU(2) (as per Example 2) describes states of the lattice that have a well defined total spin. Thus, the implementation of anyonic constraints in the MPS is closely related to and generalizes, the implementation of global onsite symmetry constraints in the MPS, as discussed in (for example) Ref.~\onlinecite{Singh101}.

\subsection{Auxiliary charges\label{ssec:supplementary}}
In Secs.~\ref{sec:one} and \ref{ssec:heisenberg} we illustrated our anyonic MPS formalism in the context of a lattice system made up of sites that are populated by a single type of anyon. However, our formalism can also be applied to the case where lattice sites contain supplementary degrees of freedom that may or may not be described by an anyon model. For instance, consider a lattice where each site $i$ corresponds to an anyon $a_i$ and a $d-$dimensional spin. The anyonic MPS formalism can be extended for such lattice systems in a rather straightforward way by treating the spin degree of freedom on each site as the degeneracy of the anyon $a_i$. That is, each site of the lattice is described by a basis $\{\ket{a_i,t_{a_i}}\}$ where the degeneracy index $t_{a_i}$ takes values $t_{a_i}=1,2,\ldots,d$. By making this identification, the anyonic MPS formalism as described in this paper can be applied to study anyon $\times$ spin lattice systems.

More generally, the lattice sites may contain supplementary degrees of freedom that correspond to charges described by a different anyon model or a global symmetry group $\mathcal{G}$. In this case each site $i$ is described by a basis $\{\ket{a_i,n_{i}, t_{a_i,n_i}}\}$ where $a_i$ is an anyon charge, $n_i$ is another anyon charge (or a symmetry charge) and $t_{a_i,n_i}$ is a degeneracy index. Our formalism can be applied to this scenario by treating the pair $(a_i, n_i)$ as a \textit{composite} charge. The fusion rules and multiplicity tensor $N^c_{ab}$, the $R$-coefficients, and the $F$-move coefficients for the composite charges $(a,n)$ can be derived from the corresponding data for the individual charges $a$ and charge $n$:
\begin{align}
N_{a_1n_1,a_2n_2}^{an} &= N_{a_1a_2}^{a}N_{n_1n_2}^{n},\\
R^{a_1n_1,a_2n_2}_{an} &= R^{a_1a_2}_{a}R^{n_1n_2}_{n},
\end{align}
\begin{align}
(F^{a_1n_1,a_2n_2,a_3n_3}_{an})&_{a_{12}n_{12},a_{23}n_{23}}\\
 &= (F^{a_1a_2a_3}_a)_{a_{12}a_{23}}(F^{n_1n_2n_3}_n)_{n_{12}n_{23}}.\nonumber
\end{align}
An application of the anyonic MPS to such a scenario is illustrated in Sec.~\ref{ssec:hubbard} in the context of the anyonic Hubbard lattice model where each site is described by a composite charge $(a,n)$; $a$ is an anyon charge and $n$ is a U(1) charge associated with the number of mobile anyons on the site.

\subsection{Fusion multiplicities\label{ssec:fusionmult}}
In certain anyon models, multiplicities $N_{ab}^c$ of fusion outcomes [\eref{eq:fusionrules}] can be greater than 1. Non-trivial multiplicities can be accomodated in the anyonic MPS formalism simply by appending a multiplicity label $\nu$ to the anyon charges output by a fusion. The $R$ and $F$ coefficients are then augmented by a multiplicity index,
\begin{equation}
\begin{split}
R^{ab}_c &\mapsto R^{ab}_{c\nu_c}, \\
(F^{abc}_d)_{ef} &\mapsto (F^{abc}_{d\nu_d})_{e\nu_e f\nu_f}.
\end{split}\label{eq:multmodels}
\end{equation}
(Strictly, charges $a$, $b$, and $c$ are also supplemented by a multiplicity index but the values of the $R$ and $F$ tensors are independent of the values of these extra indices.)

The treatment of fusion multiplicities is very similar to the treatment of \textit{degeneracies} of anyonic charges discussed above in the context of the MPS formalism. However, whereas the degeneracy index discussed above enumerates the {different} labellings of (a portion of) the fusion tree which yield the same total charge, the multiplicity index enumerates multiple copies of an output charge generated by a {single} labelling of the fusion tree. The most general anyonic tensor therefore carries three labels on each leg corresponding to charge, multiplicity, and degeneracy. The multiplicity index has been suppressed throughout this paper, and should not be confused with the number index appearing in \eref{eq:mydesc} which is an additional index specific to that model, and specifies the charge of an auxiliary U(1) symmetry group corresponding to the number of particles present at a lattice site.



\begin{thebibliography}{74}
\bibitem{Pachos}  J.K. Pachos, \textit{Introduction to Topological Quantum Computation}, Cambridge University Press (2012).
\bibitem{Haldane} F. D. M. Haldane. Phys. Lett. \textbf{93A}, 464, (1983); F. D. M. Haldane. Phys. Rev. Lett. \textbf{50}, 1153, (1983).
\bibitem{TTWL} S. Trebst, M. Troyer, Z. Wang, and A.W.W. Ludwig, Progress of Theoretical Physics Supplement No. \textbf{176}, 384 (2008).

\bibitem{Feiguin07} A. Feiguin, S. Trebst, A. W. W. Ludwig, et al., Phys. Rev. Lett. \textbf{98}, 160409 (2007).
\bibitem{FRBM} L. Fidkowski, G. Refael, N.E. Bonesteel, and J.E. Moore, Phys. Rev. B {\bf 78}, 224204 (2008).
\bibitem{Poilblanc11} D. Poilblanc, A. W.W. Ludwig, S. Trebst and M. Troyer Phys. Rev. B \textbf{83}, 134439 (2011).
\bibitem{Gils13} C. Gils, E. Ardonne, S. Trebst et.al., Phys. Rev. B \textbf{87}, 235120 (2013).
\bibitem{Poilblanc13} D. Poilblanc, A. Feiguin, M. Troyer, et. al, Phys. Rev. B \textbf{87}, 085106 (2013).

\bibitem{Fannes92} M. Fannes, B. Nachtergaele, and R. Werner, Commun.Math. Phys. \textbf{144}, 443 (1992).
\bibitem{Ostlund95} S. Ostlund and S. Rommer, Phys. Rev. Lett. \textbf{75}, 3537 (1995).
\bibitem{White92} S. R. White, Phys. Rev. Lett. \textbf{69}, 2863 (1992).
\bibitem{PerezGarcia07} D. PerezGarcia, F. Verstraete, M. M. Wolf, and J. I. Cirac, Quantum Inf. Comput. \textbf{7}, 401 (2007).

\bibitem{Vidal03} G. Vidal, Phys. Rev. Lett. \textbf{91}, 147902 (2003).
\bibitem{Vidal04} G. Vidal, Phys. Rev. Lett. \textbf{93}, 040502 (2004).
\bibitem{Daley04} A. J. Daley, C. Kollath, U. Schollwock, and G. Vidal, J. Stat. Mech. Theor. Exp., P04005 (2004).
\bibitem{White04} S. R. White and A. E. Feiguin, Phys. Rev. Lett., \textbf{93}, 076401 (2004).
\bibitem{Schollwock05} U. Schollwock, J. Phys. Soc. Jpn.,\textbf{74S}, 246 (2005).
\bibitem{Vidal071} G. Vidal, Phys. Rev. Lett., \textbf{98}, 070201 (2007).
\bibitem{Shi06} Y. Shi, L.-M. Duan and G. Vidal, Phys. Rev. A, \textbf{74}, 022320 (2006).

\bibitem{onlynonabel} If charges $\{a_i\}$ are Abelian and fixed, charges $\{u_j\}$ that appear on the internal edges of the fusion tree are completely determined by the charges $\{a_i\}$.

\bibitem{TB} H. Tran and N.E. Bonesteel, Comp. Mat. Sci. \textbf{49}, S395 (2010).

\bibitem{Vidal072} G. Vidal, Phys. Rev. Lett., \textbf{99}, 220405 (2007).
\bibitem{Vidal08} G. Vidal, Phys. Rev. Lett., \textbf{101}, 110501 (2008).
\bibitem{Trebst08} S. Trebst, M. Troyer, Z. Wang, A. W. W. Ludwig, Prog. Theo. Phys. Supp. \textbf{176}, 384 (2008).
\bibitem{Pfeifer10} R. N. C. Pfeifer, P. Corboz, O. Buerschaper, et al., Phys. Rev. B \textbf{82}, 115126 (2010).
\bibitem{Koenig10} R. Koenig and E. Bilgin Phys. Rev. B \textbf{82}, 125118 (2010).

\bibitem{ZLSPB} V. Zatloukal, L. Lehman, S. Singh, J.K. Pachos, and G.K. Brennen, arXiv:1207.500.
\bibitem{LZBPW} L. Lehman, V. Zatloukal, G.K. Brennen, J.K. Pachos, and Z. Wang, Phys. Rev. Lett. \textbf{106}, 230404 (2011).
\bibitem{Lehman2013} L. Lehman, D. Ellinas, and G.K. Brennen, Journal of Computational and Theoretical Nanoscience, \textbf{10}, 1634-1643 (2013).

\bibitem{BondersonThesis} P. H. Bonderson, \textit{Non-Abelian Anyons and Interferometry}, Ph.D. thesis, California Institute of Technology (2007).
\bibitem{kitaev2006} A. Kitaev, Ann. Phys. \textbf{321}, 2 (2006).

\bibitem{CBD} The canonical bipartite decomposition of a state $\ket{\Psi}$ of anyonic lattice $\mathcal{L}$ with total charge $a_\mathrm{tot}$ that may be different from the vacuum can be constructed as follows. Consider the lattice $\mathcal{L}'$ that is obtained by attaching a ``virtual'' anyon with charge $a^*_\mathrm{tot}$ to the right of $\mathcal{L}$. Consider state $\ket{\Psi'}$ on $\mathcal{L}'$ that has total vacuum charge and that is obtained from $\ket{\Psi}$ as\\
\begin{equation}
\Psi'_{\alpha\beta'} =  \sum_{\beta}(F^{abb^*}_{b'})_{a_\mathrm{tot}1}(\tilde{N}^{\tiny \mbox{[in]}})_{\beta a^*_\mathrm{tot}}^{\beta'} N_{a_\mathrm{tot} a^*_\mathrm{tot}}^1\Psi_{\alpha\beta},\nonumber
\end{equation}
where we have fused indices $\beta$ and $a^*_\mathrm{tot}$ into a total index $\beta'$ to obtain a bipartite decomposition of $\ket{\Psi'}$. Next, construct the Schmidt decomposition $(\Gamma^{^{[\mathcal{A}]}}, \lambda,\Gamma^{[\mathcal{B}]'})$ of state $\Psi'_{\alpha\beta'}$ as described in Sec.~\ref{ssec:bidecom}. The canonical bipartite decomposition of state $\ket{\Psi}$ is given by tensors $(\Gamma^{^{[\mathcal{A}]}}, \lambda, \Gamma^{^{[\mathcal{B}]}})$ where tensor $\Gamma^{^{[\mathcal{B}]}}$ is obtained from tensor $\Gamma^{[\mathcal{B}]'}$ as
\begin{equation}
\Gamma^{[\mathcal{B}]a_\mathrm{tot}}_{\mu \beta} = \sum_{\beta'} (F^{bb^*a_\mathrm{tot}}_{a_\mathrm{tot}})_{1b'} N_{a_\mathrm{tot} a^*_\mathrm{tot}}^1 (\tilde{N}^{\tiny \mbox{[out]}})_{\beta a^*_\mathrm{tot}}^{\beta'} \Gamma^{[\mathcal{B}]' 1}_{\mu \beta'}. \nonumber 
\end{equation}
\bibitem{hikami2008} K. Hikami, Ann. Phys. \textbf{323}, 1729 (2008).
\bibitem{pfeifer2013} R. N. C. Pfeifer, arXiv:1310.0373 [cond-mat.str-el] (2013).
\bibitem{IsingSU2} When mapping from a system of Ising anyons to SU(2)$_2$, note that the braid tensors differ. This detail is, however, unimportant for the present study as no braiding is involved in either the definition of the Hamiltonian \eeref{eq:ising} or in the implementation of the anyonic MPS update.
\bibitem{Gomez} C. Gomez, M. Ruiz-Altaba, and G. Sierra, \textit{Quantum Groups in Two Dimensional Physics}, Cambridge University Press (1996).
\bibitem{evenodd} For a $\sigma$ charge placed on each site of the lattice the bipartite decomposition of \eref{eq:aschmidt} for a partition $[1\ldots r]:[r\!+\!1 \ldots L]$ of the chain has only charges $1$ and $\psi$ when $r$ is even, and only charge $\sigma$ when $r$ is odd.
\bibitem{averagecorr} For a translationally invariant Hamiltonian, the infinite TEBD algorithm produces an MPS approximation to the ground state that is only invariant under translations by two sites. That is, the ground state is described by an infinite MPS that consists of repeating tensors $\lambda^{[A]}, \Gamma^{[A]}, \lambda^{[B]}, \Gamma^{[B]}$. The ground state energies per site listed in Table~\ref{table:gsenergy} correspond to the averages over even ($A-B$) and odd ($B-A$) bonds of the MPS.
\bibitem{restrictHop} This restriction could be relaxed without significant obstacles provided an infinite contact term is present to prohibit two anyons occupying the same position on the ladder.
\bibitem{OBC} For open boundary conditions, the 2-site terms at the boundaries differ slightly from the terms in the bulk, with the matrix elements in \eref{eq:hamnew} corresponding to tunneling on sites $1$ and $L$ being equal to $\kappa$ ($\kappa^*$) instead of $\kappa/2$ ($\kappa^*/2$).
\bibitem{orderislands} The $r$ pinned anyons in an island may be linearly ordered in an arbitrary way to define the action of the $r-$strand braid group.
\bibitem{onsite} An onsite global symmetry means that the symmetry group acts identically on each site of the lattice.
\bibitem{Singh101} S. Singh, H.-Q. Zhou, and G. Vidal, New J. Phys. \textbf{12} 033029 (2010).
\bibitem{Singh102} S. Singh, R. N. C. Pfeifer and G. Vidal, Phys. Rev. A \textbf{82}, 050301 (2010).
\bibitem{Singh11} S. Singh, R. N. C. Pfeifer and G. Vidal, Phys. Rev. B \textbf{83}, 115125 (2011).
\bibitem{Singh12} S. Singh and G. Vidal, Phys. Rev. B \textbf{86}, 195114 (2012).

\end{thebibliography}
\end{document}